\documentclass[sigconf,screen,authorversion,nonacm]{acmart}
\pdfoutput=1

\usepackage{multirow}
\usepackage{multicol}
\usepackage{subcaption}

\usepackage{pifont}% http://ctan.org/pkg/pifont
\newcommand{\cmark}{\ding{51}}%
\newcommand{\xmark}{\ding{55}}%
\usepackage{booktabs}
\usepackage{graphicx}
\usepackage{enumitem}
\usepackage{bm}
\usepackage{xcolor}
\usepackage{algorithm}
\usepackage{algpseudocode}
\usepackage{array}
\usepackage{threeparttable}
\usepackage{xparse}

\begin{document}
\title{Terabyte-Scale Analytics in the Blink of an Eye}

\author{Bowen Wu}
\authornote{Work done during an internship at the Microsoft Gray Systems Lab.}
\authornote{Equal contribution.}
\affiliation{\institution{ETH Zurich}\country{}}
\email{bowen.wu@inf.ethz.ch}

\author{Wei Cui}
\authornotemark[2]
\affiliation{\institution{Microsoft Research Asia}\country{}}
\email{weicu@microsoft.com}

\author{Carlo Curino}
\affiliation{\institution{Gray Systems Lab, Microsoft}\country{}}
\email{carlo.curino@microsoft.com}

\author{Matteo Interlandi}
\affiliation{\institution{Gray Systems Lab, Microsoft}\country{}}
\email{matteo.interlandi@microsoft.com}

\author{Rathijit Sen}
\affiliation{\institution{Gray Systems Lab, Microsoft}\country{}}
\email{rathijit.sen@microsoft.com}

\begin{abstract}
For the past two decades, the DB community has devoted substantial research to take advantage of cheap clusters of machines for distributed data analytics---we believe that we are at the beginning of a paradigm shift.
The scaling laws and popularity of AI models lead to the deployment of incredibly powerful GPU clusters in commercial data centers. Compared to CPU-only solutions, these clusters deliver impressive improvements in per-node compute, memory bandwidth, and inter-node interconnect performance.  

In this paper, we study the problem of scaling analytical SQL queries on distributed clusters of GPUs, with the stated goal of establishing an upper bound on the likely performance gains. To do so, we build a prototype designed to maximize performance by leveraging ML/HPC best practices, such as group communication primitives for cross-device data movements.
This allows us to conduct thorough performance experimentation to point our community towards a massive performance opportunity of at least 60$\times$. 
To make these gains more relatable, before you can blink twice, our system can run all 22 queries of TPC-H at a 1TB scale factor!
\end{abstract}

\maketitle

\pagestyle{plain}

\section{Introduction}
\label{sec:intro}

The massive parallelism and multi-TB/sec memory bandwidths offered by modern GPUs are hugely beneficial for accelerating SQL analytics queries. Consequently, GPU acceleration for SQL analytics continues to receive much attention, and numerous recent studies have extended the state of the art to accelerate query processing on GPUs~\cite{surakav_he_2022,crystal_shanbhag_2020,Yogatama25-lancelot,mohr23-boss}. 
At the same time, driven by the enormous potential and need for supporting the training and inferencing of GenAI models, GPU technology has been rapidly evolving, both within the GPU hardware itself, as well as at the system level with the introduction of multi-GPU machines and high-bandwidth interconnects~\cite{h100,mi300x,nvl72}.  
We believe that such advances have opened new opportunities for SQL acceleration, at scales and speeds unimaginable just a few years ago. Multi-GPU machines will drive the next wave of acceleration for analytics. 

Single-GPU acceleration for SQL analytics has traditionally faced two performance bottlenecks~\cite{Rosenfeld22-revisit}: (1) limited high-bandwidth memory (HBM) capacity on the GPU; and (2) low CPU-GPU data movement bandwidth (over PCIe) compared to available CPU-memory bandwidths. 
The recent availability of multi-GPU machines from NVIDIA and AMD has alleviated these concerns by offering opportunities for resource aggregation. For example, the NVIDIA A100/H100 and AMD MI300X eight-GPU machines have an aggregate HBM capacity of 640 GiB and 1.5 TiB per machine, respectively, thereby enabling larger datasets to remain resident in GPU HBM without requiring frequent CPU-GPU transfers. With a separate PCIe link per GPU, the aggregate CPU-GPU PCIe bandwidth in multi-GPU machines is comparable to the CPU main memory bandwidths available in high-end dual-socket servers today. 
Additionally, multi-GPU machines support high-bandwidth data exchange between GPUs, within and across machines, thereby enabling high scale-up and scale-out performance for distributed query processing---data exchange operations for shuffle and broadcast play a critical role in running analytical queries at scale and are often limited by the network bandwidth. But not anymore!

\begin{figure}[t]
\centering
\subfloat[Architecture]{\includegraphics[width=0.56\columnwidth]{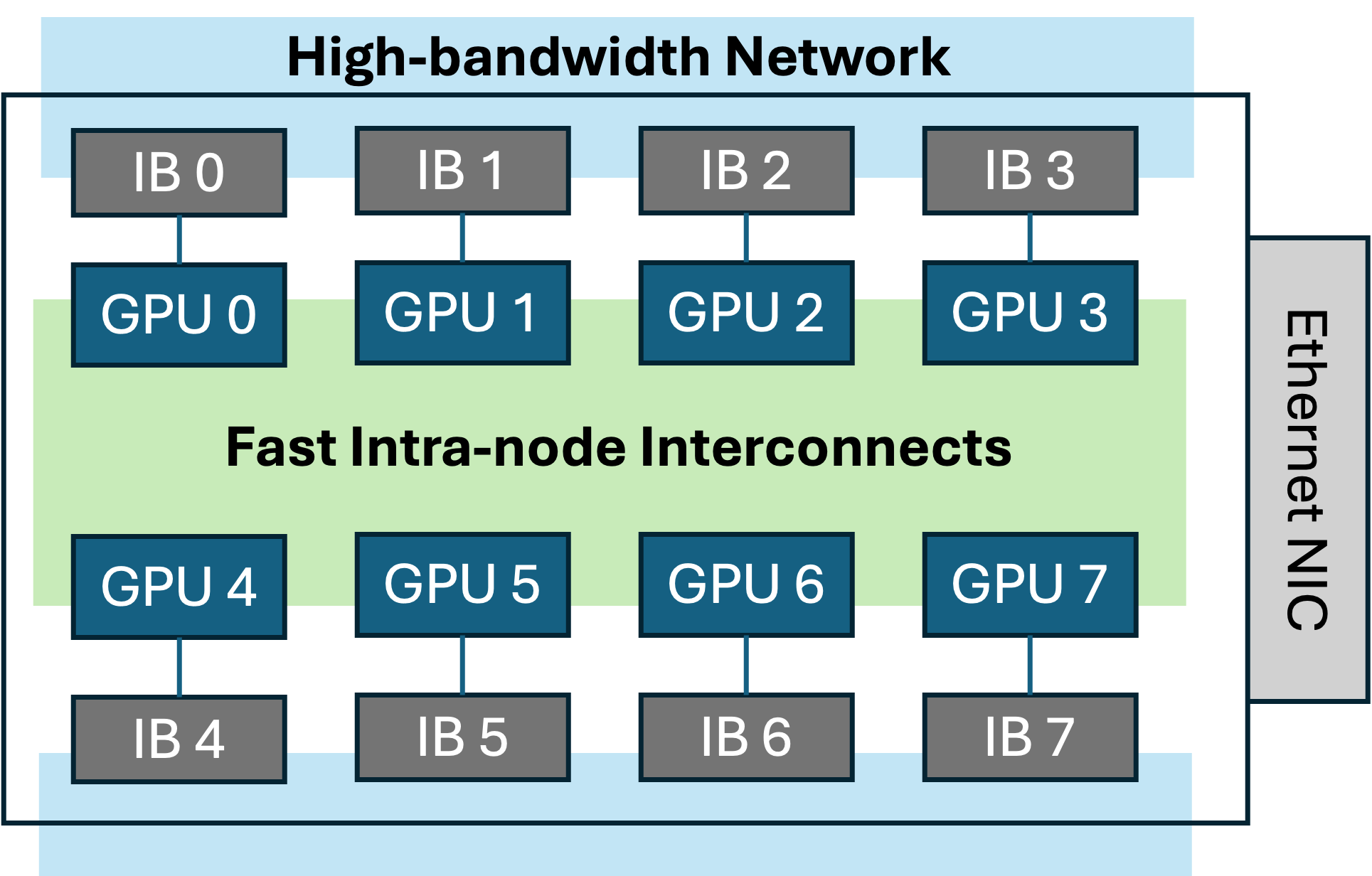}\label{fig:multigpu-arch}\vspace{-5pt}}\hspace{15pt}
\subfloat[Best times]{\includegraphics[width=0.2\columnwidth]{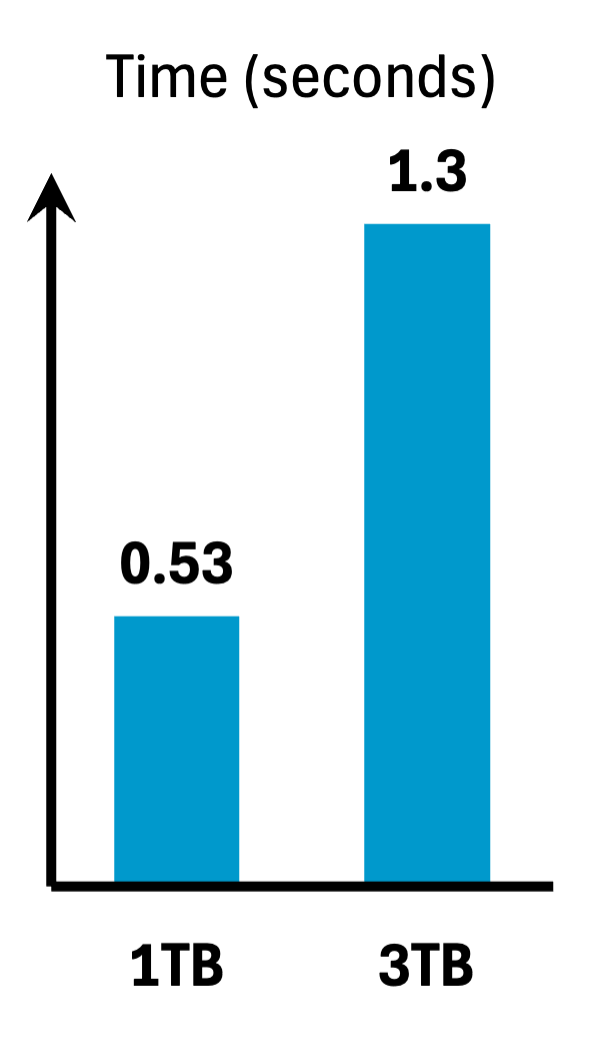}\label{fig:tpch-best-times}\vspace{-5pt}}
\vspace{-8pt}
\caption{(a) Multi-GPU machine architecture. IB: InfiniBand. (b) Best total run time for all 22 queries of TPC-H using 5 machines (40 GPUs).
\vspace{-10pt}}
\label{fig:VM-arch}
\vspace{-2ex}
\end{figure}
Figure~\ref{fig:multigpu-arch} shows the architecture of a scale-up 8-GPU machine, which we use as a building block for our scale-out cluster setups. The GPUs are interconnected by a high-bandwidth backplane network, made up of NVLinks and NVSwitches for NVIDIA multi-GPU machines, and Infinity Fabric for AMD multi-GPU machines. The bandwidth for both outgoing and incoming traffic to each GPU is in the order of several hundred GB/sec (e.g., up to 450 GB/sec on the NVIDIA 8-H100 DGX machines, which is more than 7$\times$ the CPU-GPU PCIe gen5 bandwidth), and also exceeds the single-socket CPU-main memory bandwidths of most modern servers. The inter-GPU backplane network within these machines offers up to 3.6 TB/sec of aggregate bandwidth for shuffle operations, far exceeding what is available across NUMA CPU sockets~\cite{li2013numa}.
These multi-GPU machines also have impressive scale-out performance. Our highest-performing clusters provide up to 400 GB/sec bandwidth between machines with eight Mellanox NICs, one per GPU. As we will show, this can speed up SQL query performance by more than an order of magnitude compared to using traditional lower-bandwidth Ethernet connectivity (100 Gbits/sec) between machines.   
Finally, to make things practical,
such machines are available in the cloud today~\cite{aws-h100-instance,gcp-gpu-instance,azure-vm-pricing,vastai-gpu,lambda-gpu}, thereby expanding their access and affordability to the community with on-demand pricing models. 

In this work, we show that distributed SQL query processing at very high speeds is possible using a cluster of multi-GPU machines with high-bandwidth interconnects. 
While prior systems, such as HeavyDB
and Theseus~\cite{voltron-data} have explored SQL acceleration with multiple GPUs, in this paper, we push the boundary and show what is the art of the possible for SQL analytics on powerful clusters of GPU machines.
To achieve this, we implemented a distributed version of Tensor Query Processor (TQP)~\cite{surakav_he_2022}. 
With TQP, we took the nonconventional decision of leveraging high-performance ML frameworks (i.e., PyTorch) for running SQL analytics on GPUs.
The key idea was to leverage any system improvement in the ML space also for SQL analytics~\cite{Koutsoukos21-tensor}.
In this paper, we 
follow a similar philosophy by implementing data exchange operations using the core group communication primitives used in AI training. 
We leverage NVIDIA and AMD high-performance implementations of such primitives (i.e., NCCL~\cite{nccl} and RCCL~\cite{rccl}), and natively 
use their proprietary backend networks, as well as leveraging their algorithms for efficient multi-GPU cross-machine data transfers. 
Starting from input data partitioned and loaded in GPU HBMs, we run all 22 TPC-H~\cite{tpch} queries at 1 TB scale in 0.53 seconds in total using 40 GPUs (H100, 5 machines) and 3 TB scale in 1.3 seconds (Figure~\ref{fig:tpch-best-times}). We can also complete all 22 queries at 1 TB scale in 1.06 seconds on a single machine with 8 GPUs (MI300X). This beats custom scale-up CPU machines fitting the workload in RAM by over 60$\times$~\cite{HPE_TPC-H_2022,HPE_TPC-H_3TB_2024} and more for commodity scale-out cloud-based hardware.
This is just a snapshot in time of the achievable performance with the hardware that we have access to. In fact, and as we write, next-generation GPUs with even faster network interconnect are being deployed in the cloud~\cite{blackwell}. To address how future-generation hardware and interconnects can impact end-to-end performance, we also developed analytical models to get insights into the expected performance as we continue to scale and run TQP on new hardware. We expect the performance that we can achieve on GPU clusters with fast interconnects to drastically outpace CPU equivalents.

In summary, this paper makes the following contributions:
\begin{enumerate}[wide, labelwidth=!, labelindent=0pt, topsep=0pt]
\item We show how off-the-shelf group communications libraries that are developed primarily for AI applications can be adapted to be applied to the SQL analytics domain.
\item We show how TQP can be easily extended to take advantage of multiple GPUs across a cluster of machines coming from different vendors, thereby enjoying ease of portability along with highly competitive scale-up and scale-out performance.
\item We demonstrate, for the first time, TPC-H 1 TB total query performance of less than a second, and < 1.5 seconds at a 3 TB scale using a cluster of multi-GPU machines in the cloud. 
\item To gain insights about scalability and extrapolate performance to future hardware, we present analytical performance models of the time-consuming shuffle and broadcast operations. 
\item We characterize the execution of TPC-H workloads on a multi-GPU cluster from multiple angles and analyze the effect of various factors on workload performance, such as warm/cold run, broadcast implementation, data skew, and data placement. The analysis gives valuable insights into designing an efficient SQL analytics system for a multi-GPU cluster.
\end{enumerate}

We believe that this paper firmly demonstrates the potential for accelerating analytics at scale over multi-GPU clusters in the cloud.  

The rest of the paper is organized as follows. We describe our distributed SQL query acceleration approach using the Tensor Query Processor (TQP) and off-the-shelf collective communications libraries in Section~\ref{sec:approach}. We present analytical performance models for data exchange operations in Section~\ref{sec:models}, with experimental evidence of scaling trends in Section~\ref{sec:data-exchange-analysis}. Section~\ref{sec:experimental-setup} lists our cluster configurations. Sections~\ref{sec:tpch-analysis} and~\ref{sec:perf-sensitivity} analyze terabyte-scale TPC-H performance on standard (uniform) and skewed data, respectively, and explore other factors contributing to the performance.
Section~\ref{sec:related} summarizes related literature and Section~\ref{sec:conclude} concludes the paper.

\section{Approach: Distributed TQP}
\label{sec:approach}

We extend and use TQP to accelerate SQL queries using multiple GPUs. Our approach is to add data exchange operators to tensor programs generated by TQP and use primitives provided by the NVIDIA Collective Communications Library (NCCL) API to implement them. We describe each of these components in this section. 

\subsection{Background: Tensor Query Processor}
\label{sec:approach-tqp}

Tensor Query Processor (TQP)~\cite{surakav_he_2022, tensor_tea_vldb_2022} accelerates analytical queries by implementing relational operators using PyTorch's tensor API. By leveraging PyTorch, TQP can execute queries on specialized hardware such as GPUs~\cite{surakav_he_2022,gandhi2022tensor} and APUs~\cite{tqp-xbox:damon:2023}. The advantage of using PyTorch's tensor API not only relies on the ease of portability to diverse and rapidly evolving hardware platforms but also on taking advantage of any algorithm improvement coming from the ML community, while maintaining a familiar programming interface. 

TQP is composed of a set of utilities for (1) converting input data into tensor format, and (2) a query compiler transforming input queries into tensor programs. 
For the former, TQP supports Pandas DataFrame~\cite{pandas}, Parquet~\cite{parquet}, Numpy~\cite{numpy}, and CSV as input formats. TQP also supports integers, floating point as well as ASCII string (in dictionary and value-encoded formats), and date inputs\footnote{Notably, decimals are not supported yet. Decimal columns are currently loaded as floating-point numbers.}.
For the latter, given an input query, TQP uses Apache Spark's~\cite{spark} query optimizer (Catalyst) to generate an initial physical query plan for single-node execution. TQP then tensorizes~\cite{surakav_he_2022} the plan, and compiles it into a tensor program that is fed with the input data (in tensor format) to generate the query results. 
We reuse TQP's compilation stack in this work but add additional capabilities for supporting distributed SQL query processing, as we describe next. 

\subsection{Background: Data Exchange for Distributed Query Processing}
\label{sec:distributed-tqp}

Data exchange is a costly operation because it happens over networks having maximum bandwidths that can be an order of magnitude (or more) lower than device-local HBM bandwidths. 
Thus, distributed analytical systems strive to minimize data exchanges and maximize local computations. For example, queries that operate on single tables (e.g., TPC-H Q1, Q6) do not need to exchange data, except for a final aggregation step. An optimal partitioning of the input tables can also avoid data exchanges for some queries (e.g., TPC-H Q12) even though they operate on multiple tables. All other queries would need to exchange data during execution. Shuffle and broadcast are the most common kinds of data exchange operations.

\begin{figure}[ht]
\centering
\includegraphics[width=0.95\columnwidth]{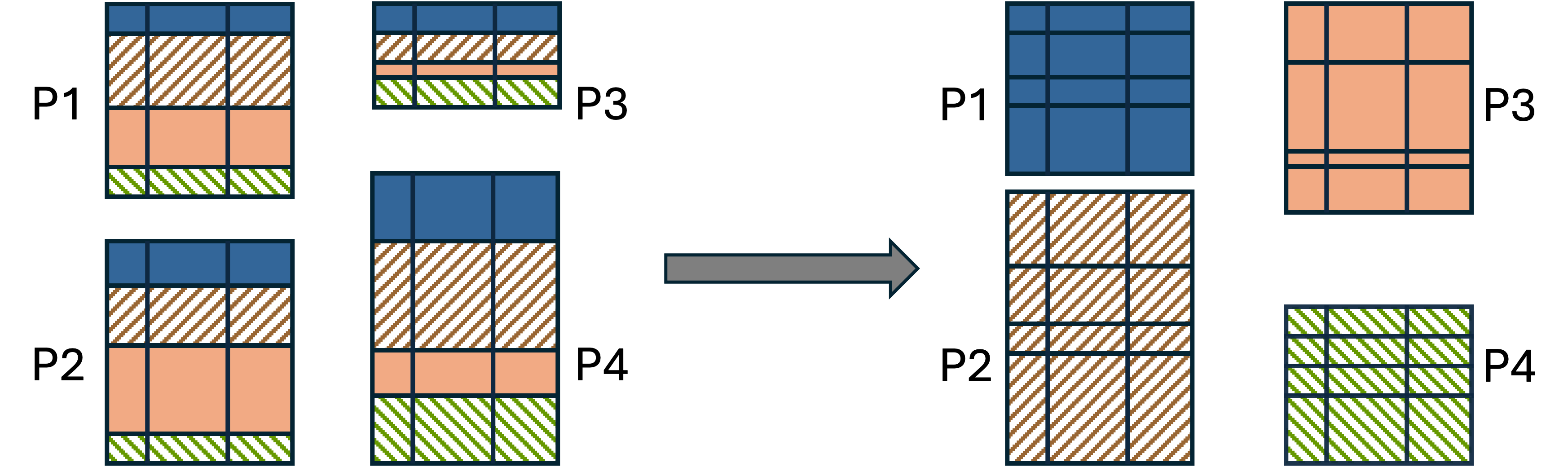}
\caption{Example shuffle operation between 4 processes, P1--P4, on a distributed table with 3 columns.\vspace{-5pt}}
\label{fig:shuffle-example}
\end{figure}
Figure~\ref{fig:shuffle-example} shows an example of a shuffle operation between 4 processes, P1--P4. The table with 3 columns, in this example, is distributed in each of the 4 processes, with the parts not necessarily being of the same size (e.g., data skew). 
Each process then applies the same partitioning function (based on one or more keys/attributes of the table) to partition its data. 
The shuffle operation then redistributes the partitions among the processes so that each process has all partitions corresponding to the same set of key values.  

\begin{figure}[ht]
\centering
\includegraphics[width=0.9\columnwidth]
{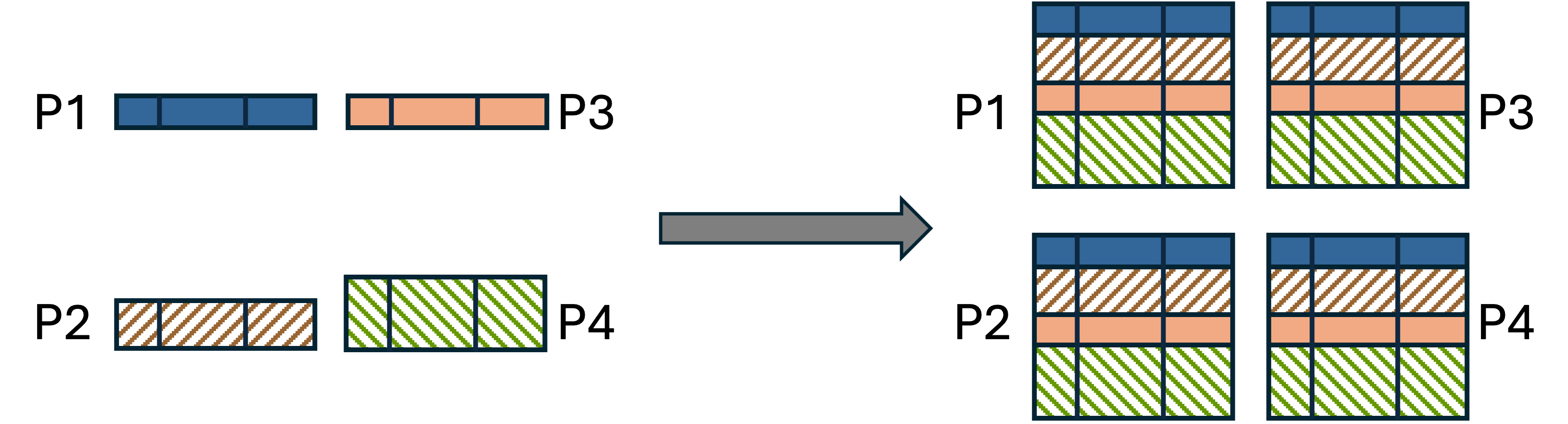}
\vspace{-1ex}
\caption{Example broadcast operation between 4 processes, P1--P4, on a distributed table with 3 columns.\vspace{-5pt}}
\label{fig:broadcast-example}
\end{figure}
Figure~\ref{fig:broadcast-example} shows an example of a broadcast operation. Here, each process scatters its part of the distributed table to all other processes. At the end of the operation, the entire table is replicated in all processes. While shuffle redistributes data while keeping the total data across all participants fixed, broadcast replicates data, thereby increasing the total data size across all participants. Broadcasts are useful for small tables since they can enable local joins without needing the other table to be partitioned on the join keys.

Generating optimal query plans for distributed execution requires knowledge of how the data is partitioned across the processes~\cite{databricks,spark,bruno24-uqo,bigquery}. For example, a join between two partitioned tables could proceed in one of the following ways:
\begin{itemize}[leftmargin=*]
    \item If both tables are partitioned on the join keys, then join locally.
    \item Broadcast one of the tables and then join locally.
    \item Partition and shuffle both tables and then join locally.
\end{itemize}
 
\begin{figure}[h]
\vspace{-1ex}
    \centering
    \includegraphics[width=\columnwidth]{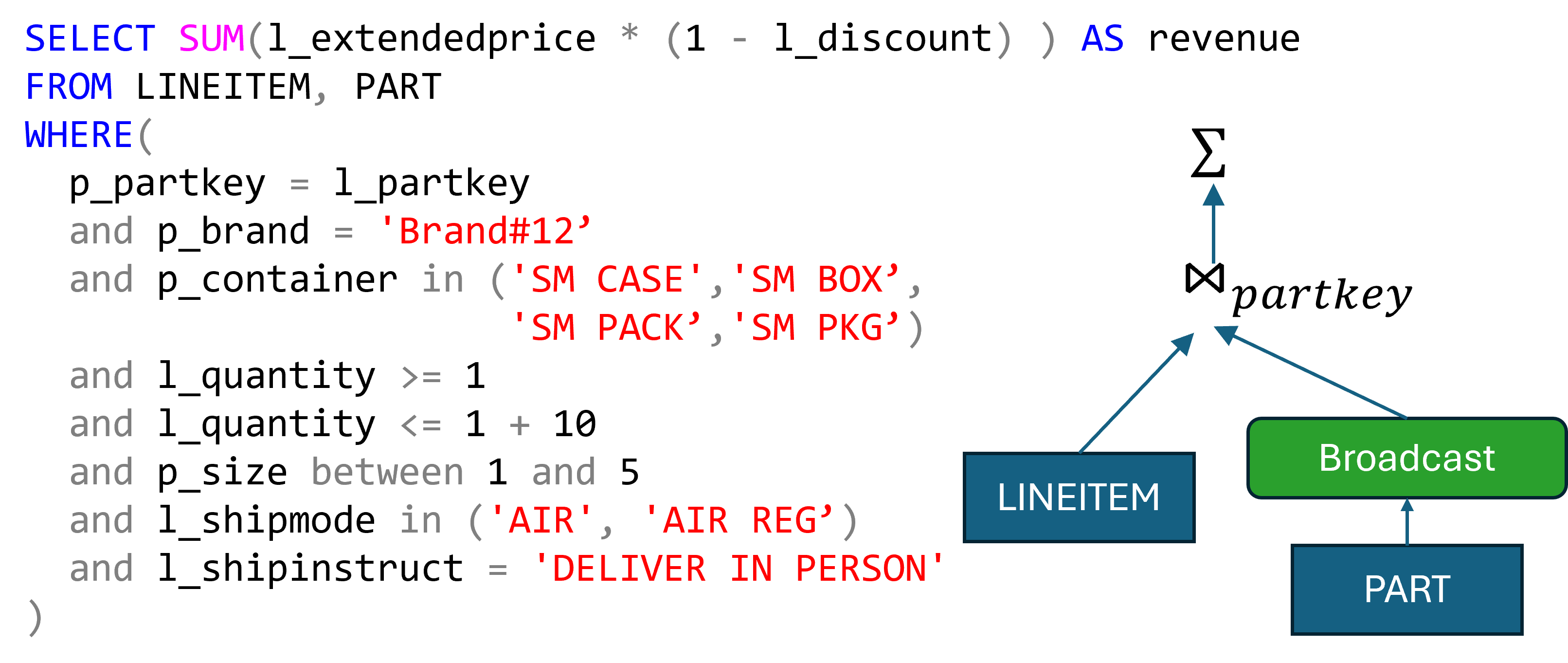}
    \vspace{-20pt}
    \caption{Example query and simplified plan.\vspace{-5pt}}
    \label{fig:example-plan}
\end{figure}

Figure~\ref {fig:example-plan} shows an example using a part of TPC-H Q19 and its query plan. The input tables to the join, \texttt{LINEITEM} and \texttt{PART} (both after applying filter predicates), are not partitioned by their join keys (\texttt{l\_partkey}, \texttt{p\_partkey}), and hence the plan requires a data exchange before the join operation. One option would be to shuffle both \texttt{LINEITEM} and \texttt{PART} so that they get partitioned using their join keys. Another option is to broadcast the smaller table, \texttt{PART} (filtered), instead. This second option does not require a (costly) shuffle of \texttt{LINEITEM} (filtered) before the join can proceed, but uses up more memory to hold the broadcasted \texttt{PART} (filtered) table. In this case, we insert the broadcast operator in the query plan as shown in the figure.

\subsection{Collective Communications Library}
\label{sec:approach-nccl}

The open-source NVIDIA collective communication library (NCCL) provides a variety of communication primitives between multiple GPUs. These include collective communication involving all participating GPUs (e.g., all-reduce) as well as point-to-point (p2p) communication for send and receive operations. We use NCCL primitives to implement the data exchange operations in TQP. This approach has the following advantages.
\begin{itemize}[leftmargin=*]
    \item NCCL transparently handles various available interconnect technologies for GPU-GPU communications, 
    such as NVLink, GPUDirect RDMA over InfiniBand (IB), TCP/IP over Ethernet, etc.
    These interconnects can be either intra-node (e.g., NVLink connects GPUs local to a VM) or inter-nodes (e.g., RDMA over IB connects remote GPUs not residing on the same node).
    \item NCCL optimizes data routing for the network topology and chooses different routing algorithms for the underlying platform as part of its bootstrapping process.
    \item NCCL's APIs are supported by both NVIDIA and AMD. The AMD implementation is named RCCL (ROCm Communication Collectives Library)~\cite{rccl}, but the APIs and semantics remain the same. Thus, adopting NCCL enables ease of portability across vendors, along with enjoying vendor-maintained improvements for new hardware generations, which is similar to TQP's motivation of leveraging PyTorch's API for implementing SQL operators.  
\end{itemize}
\begin{table}[ht]
    \centering
    \vspace{-2ex}
    \caption{NCCL operations used in TQP Analytics.}\vspace{-8pt}
    \begin{tabular}{|c|c|}\hline
    \texttt{\textbf{ncclSend}}& Send for point-to-point transfer\\\hline
    \texttt{\textbf{ncclRecv}}& Receive for point-to-point transfer\\\hline
    \texttt{\textbf{ncclGroupStart}}& Start of a group of operations\\\hline
    \texttt{\textbf{ncclGroupEnd}}& End of a group of operations\\\hline
    \texttt{\textbf{ncclAllReduce}}& Reduction with aggregate operation\\\hline        
    \texttt{\textbf{ncclBroadcast}}& Point-to-all Broadcast operation\\\hline
    \end{tabular}
    \label{tab:nccl-calls}
    \vspace{-1ex}
\end{table}

Table~\ref{tab:nccl-calls} shows the set of NCCL APIs that we used for implementing data exchange operations in TQP.
\texttt{ncclAllReduce} are collective operations for scalar aggregation, whereas \texttt{ncclSend} and \texttt{ncclRecv} are p2p operations that can be used to build up data exchange operations for broadcast and shuffle. \texttt{ncclGroupStart} and \texttt{ncclGroupEnd} are used to enclose a group of individual NCCL operations to reduce the number of kernel invocations and to achieve better routing that improves bandwidth utilization.

The reason why we need to use p2p operations is due to a mismatch between NCCL collective APIs, which were developed primarily for ML applications, and what is needed to implement data exchange for distributed query execution. 
For example, collective operations
assume equal-sized data at each node (as is the case for ML workloads). For SQL queries, this is restrictive since partitioned data is not always perfectly balanced. This can happen both due to skew in data distributions and having different selectivities of filter operations at different nodes. One workaround could be to use data padding to make the data sizes equal, but that would involve a waste of GPU HBM capacity as well as network bandwidth. 

To solve this mismatch between collective primitives and distributed SQL processing, one can build them using the p2p NCCL operations \texttt{ncclSend} and \texttt{ncclRecv}. These enable the construction of more complex operations such as shuffle\footnote{Note that shuffle could be expressed as an MPI \texttt{AllToAll} operation. However, NCCL doesn't implement the \texttt{AllToAll} collective operation.}, and also allow each send-receive pair to have a different size, thereby offering more flexibility.
We implement the shuffle operation among $N$ GPUs using $N^2$ (including self-transfers) \texttt{ncclSend} and $N^2$ \texttt{ncclRecv} enclosed within a \texttt{ncclGroupStart} and \texttt{ncclGroupEnd}. Algorithm~\ref{alg:shuffle} shows the pseudo-code of the shuffle executing at each GPU.\vspace{-5pt}

\begin{minipage}{0.42\columnwidth}
\small
\begin{algorithm}[H]
\caption{Shuffle @$i$}\label{alg:shuffle}
\begin{algorithmic}[1]
\Require receiving message sizes from all $j\in N$.
\State \texttt{ncclGroupStart}
\For{$j=0$ to $N-1$}
\State \texttt{ncclSend}: $i\rightarrow j$
\State \texttt{ncclRecv}: $i\leftarrow j$
\EndFor
\State \texttt{ncclGroupEnd}
\end{algorithmic}
\end{algorithm}\vspace{1pt}
\end{minipage}\hfill
\begin{minipage}{0.42\columnwidth}
\small
\begin{algorithm}[H]
\caption{Broadcast @$i$}\label{alg:broadcast}
\begin{algorithmic}[1]
\Require receiving message sizes from all $j\in N$.
\State \texttt{ncclGroupStart}
\For{$j=0$ to $N-1$}
\State \texttt{ncclBroadcast}: recv from $j$ and send if $j==i$
\EndFor
\State \texttt{ncclGroupEnd}
\end{algorithmic}
\end{algorithm}\vspace{1pt}
\end{minipage}

For broadcast operations, a similar approach of using a set of p2p exchanges, although functionally sufficient, is not optimal in performance since that sends the same data packets to the same remote machine multiple times. 
For example, GPU 0 in machine 1 needs to send the same data twice to GPU 0 and 1 in machine 2 separately. 
Using the \emph{one-to-all} broadcast operation in the communication library, e.g., \texttt{ncclBroadcast}, on the other hand, may avoid or reduce repeated transfers of the broadcasted data across machines~\cite{mpi-coll,sanders09-twotreebroadcast}. As we will show in Section~\ref{sec:coll-vs-p2p}, the impact is severe, particularly for multi-machine deployments where the network bandwidth between machines is about an order of magnitude lower with InfiniBand and about two orders of magnitude lower with Ethernet compared to intra-machine inter-GPU (e.g., NVLink) bandwidths. Hence, we use the one-to-all \texttt{ncclBroadcast} to implement the broadcast operation. Algorithm~\ref{alg:broadcast} shows the pseudo-code of the broadcast executing at each GPU. The \texttt{for} loop is needed since NCCL uses the same operation for both sending and receiving: 
when $i==j$ GPU $i$ will send, while all GPUs participating in the same \texttt{ncclBroadcast} will be receiving from GPU $i$.

The shuffle and broadcast operations are preceded by an information exchange where the data sizes are exchanged between senders and receivers. This allows the receivers to allocate receive buffers accurately. Furthermore, since we exchange one column at a time, knowledge of the incoming data sizes allows in-place construction of a contiguous tensor for each column, as needed by TQP for performing operations on it. We achieve this by determining a starting offset for data incoming from each sender in the contiguous receive buffer for the column at the recipient.
This information exchange is very lightweight compared to the shuffle or broadcast because only $N$ integers are sent from each GPU, where $N$ is the total number of GPUs participating in the data exchange. 

The reduction operation \texttt{ncclAllReduce} is useful for queries needing to do a final aggregation along with gathering data from all GPUs. The aggregation needs to be one of the supported types (currently: sum, product, minimum, maximum, and average).

\subsection{Distributed TQP}
Let's now combine everything and discuss how we implemented distributed query processing in TQP.
As a distributed processing model, we use data-parallel computing, with GPUs as the computational backends. The input data is partitioned between the GPUs in the distributed system and loaded in their device memories (HBMs). We run a TQP process for each GPU, with all the processes launched as MPI (Message Passing Interface)~\cite{mpi41} jobs by a distributed job runner (e.g., \texttt{mpirun}). 
In distributed mode, each TQP process computes using its assigned GPU on its local data, and it performs data exchange with other TQP processes, when necessary, by moving data directly between GPU HBMs wherever possible. To achieve this, we modified the TQP compiler stack to automatically inject data exchange operations (implemented as described in the previous Session) into their compiled tensor programs before group by, join operations, or for final aggregation.

Note that: (1) each TQP process executes the exact same tensor program, the only difference is that each GPU reads a different input data partition; (2) data exchange operations are managed by the processes themselves and not orchestrated by a separate ``driver'' process; (3) fault tolerance is based on re-execution (since queries run at interactive speed). 
\emph{This is closer to how ML training runs are executed, rather than how distributed analytical systems work.}
Fundamentally, we are advocating not only for leveraging ML frameworks (i.e., PyTorch and NCCL) for targeting hardware accelerators and fast network interconnect, but also that we can embrace the same computational model used for ML training.

\section{Analytical Performance Models}
\label{sec:models}
In this section, we will describe performance models that facilitate our analysis of shuffle and broadcast operations on multiple GPUs spanning one or more machines. The goal of these analyses is twofold---to gain insights about the performance scalability of data exchange operations and to extrapolate performance impact with different interconnects, larger cluster sizes, and next-generation network technologies that are not yet available.

\vspace{-1ex}
\subsection{Model Description/Assumptions} 
\label{sec:model-desc}
\begin{table}[ht]
\vspace{-2ex}
    \centering
    \caption{Notations.}\vspace{-8pt}
    \resizebox{\columnwidth}{!}{
    \begin{tabular}{|c|c|}\hline
    \textbf{k}& Number of GPUs per machine (or node, VM)\\\hline
    \textbf{V}& Number of machines (or nodes, VMs)\\\hline
    \textbf{N}& Total number of GPUs = $k\times V$\\\hline
    \bm{$B_g$}& Unidirectional inter-GPU bandwidth within each machine\\\hline
    \bm{$B_n$}& Unidirectional network bandwidth at each machine\\\hline
    \textbf{S}& Total dataset size processed by the GPU cluster. \\\hline
    \bm{$G_{ij}$}& The $i$-th GPU in the $j$-th machine.\\\hline
    \bm{$m_{ij\rightarrow pq}$}& The message sent from $G_{ij}$ to $G_{pq}$.\\\hline
    \end{tabular}
    }
    \label{tab:notations}
    \vspace{-1ex}
\end{table}

Table~\ref{tab:notations} describes the notations that we use in this paper. To begin with, we focus on the case where we exchange a large amount of data in the shuffle and broadcast operations because frequent latency-bound small-sized data transfers are not common or expensive in databases. Later, we also discuss how we adapt our models for small message sizes. Moreover, we only model the data exchange among the GPUs -- metadata
exchanges prior to the shuffle and broadcast (Section~\ref{sec:approach-nccl}) are ignored.
\begin{itemize}[leftmargin=*]
    \item The system has $V$ machines and each machines has $k$ GPUs. Machines are connected by the network (e.g., Ethernet, RDMA). 
    Each node can send and receive data to/from any other nodes.
    Each node has outbound/inbound network bandwidth $B_n$, hence each GPU has a share of $B_n/k$ network bandwidth.
    \item Within each machine, there are also pairwise connections between GPUs.  
    Each GPU can send data to all other GPUs within the same machine (i.e., local peers) at an aggregated rate of $B_g$. 
    \item The total dataset is $S$ in size. It is evenly distributed across $N$ GPUs, making the size of each GPU's work set $S/(Vk)$. Later in this section, we will model the effect of the data skew without assuming a uniform distribution.
    \item A \emph{message} is what a GPU sends to another GPU during a shuffle or broadcast operation.
    \item We assume that intra-node data transfer and inter-node data transfer can happen concurrently.
    \item We define the throughput of the operation as the total dataset size $S$ divided by the total time $T$ used to finish the operation.
\end{itemize}

\subsection{Broadcast Scalability}
\label{sec:models-broadcast}
We model the broadcast by assuming it uses a ring-based algorithm, which is true for throughput-optimized applications~\cite{hu2025demystifyingncclindepthanalysis}. To address the discrepancy in bandwidth between a GPU's intra-VM link (300-450 GB/sec) and its network (6.25-50 GB/sec), NCCL forms multiple rings, each of which connects all GPUs and spans both inter- and intra-node links. For example, in our H100+IB cluster (Table~\ref{tab:cluster_config}), every two rings share every two NICs per node, and in total 8 rings are formed. The rings become the most efficient when $B_n = B_g$, which means neither the inter- or intra-interconnect is underutilized. This approach essentially obliterates the heterogeneity of interconnects. Since the ring-based broadcast proceeds in $(N-1)$ steps, and in each step, $S/N$ data are transferred over each hop of the ring, we can calculate the broadcast time as follows.

\begin{equation*}
\vspace{-0.5ex}
    \text{T}_{\text{broadcast}} = (N-1)\frac{S/N}{\min(B_n, B_g)}\text{, for }V>1.
\end{equation*}

$S/N$ is the message size in the broadcast, which is the same as that of each GPU's work set. The throughput of the ring is bound by the minimum of inter-VM and intra-VM interconnects, i.e., $\min(B_n,B_g)$. The throughput of broadcast can be calculated as

\vspace*{-1ex}
\begin{equation}
\label{eq:broadcast-no-skew}
\boxed{
    \text{Thpt}_{\text{broadcast}} = \frac{N}{N-1}\min(B_n, B_g)\text{, for }V>1.
}
\end{equation}

\begin{figure}[t]
    \centering
    \subfloat[Broadcast.]
    {\includegraphics[trim={0ex 4ex 0ex 0ex},clip,width=0.49\columnwidth]{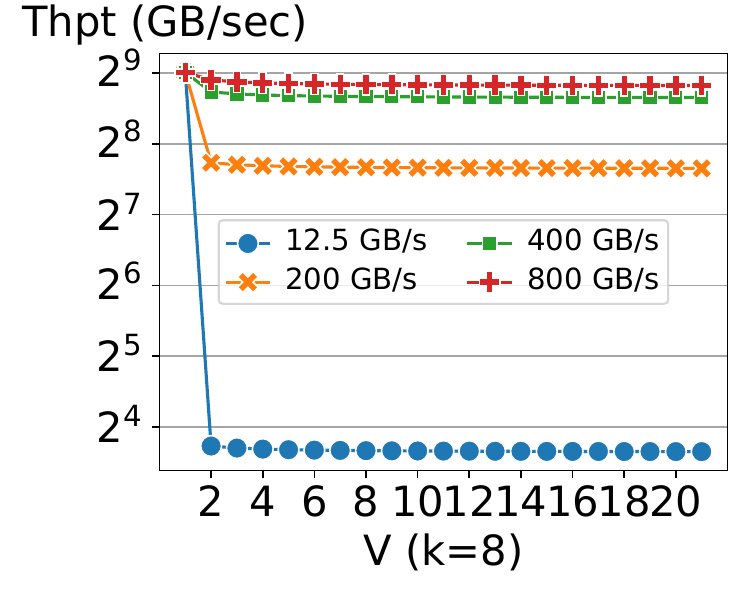}\label{fig:broadcast-model}\vspace{-1ex}}
    \subfloat[Shuffle.]
    {\includegraphics[trim={0ex 4ex 0ex 0ex},clip,width=0.49\columnwidth]{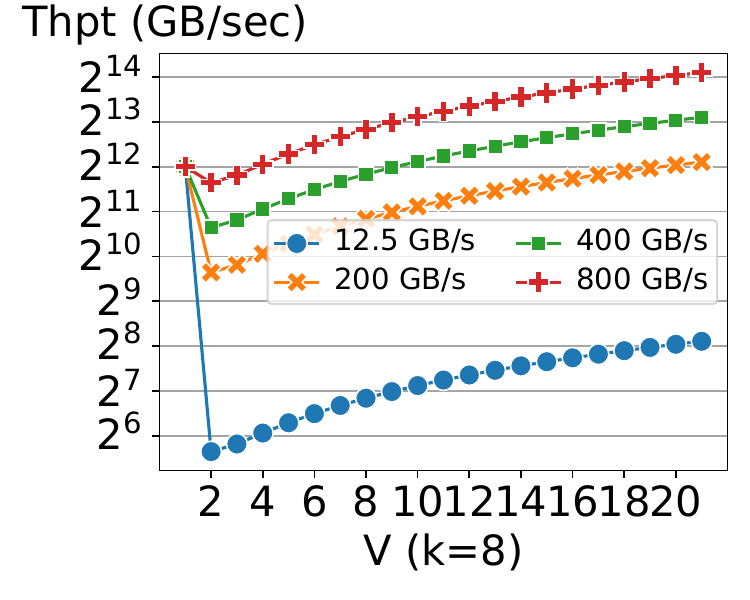}\label{fig:shuffle-model}\vspace{-1ex}}
    \vspace{-2ex}
    \caption{Model-predicted throughput}
    \label{fig:model-thpt}
 \vspace{-1ex}
\end{figure}

When $V=1$, since only the intra-VM links matter, the throughput becomes $\frac{N}{N-1}B_g$.
The above equation implies that when we add more nodes to the system ($V\uparrow$), the throughput of the broadcast will decrease. In other words, the broadcast operation does not scale with the number of nodes. Figure~\ref{fig:broadcast-model} shows the model-predicted throughput when we increase the number of nodes for different network bandwidths ($B_n$). The throughput decreases with the number of nodes and eventually converges.
Notice how 800 GB/sec network will not give much better broadcast performance than a 400 GB/sec because the intra-VM interconnects become the bottleneck.

\subsection{Shuffle Scalability}
\label{sec:models-shuffle}

The biggest difference between shuffle and broadcast is that the point-to-point message size in the shuffle operation is $S/(Vk)^2$ because
each pair of ncclSend/Recv primitives deals with only $1/(Vk)$ of the work set of a GPU whose size is $S/(Vk)$. Based on common cluster configurations, we assume that the local sends are faster than the remote sends. Therefore, the time of the shuffle operation can be calculated as
\vspace*{-1ex}
$$T=\frac{S(V-1)k^2}{(Vk)^2B_n}=\frac{S(V-1)}{V^2B_n}.$$

The throughput of the shuffle operation is therefore
\begin{equation}
\label{eq:shuffle-no-skew}
\boxed{\text{Thpt}_{\text{shuffle}}=\frac{V^2}{V-1}B_n\text{, where } V \ge 2.}
\end{equation}

When $V=1$, $\text{Thpt}_{\text{shuffle}} = \frac{N^2}{N-1}B_g$.
For $V>1$, $\text{Thpt}_{\text{shuffle}}$ is an increasing function of $V$, which means that with more nodes, the shuffle will become more efficient. 
Compared with $\text{Thpt}_{\text{broadcast}}$, shuffle is almost $V$ times more efficient than the broadcast. Figure~\ref{fig:shuffle-model} shows the model-predicted throughput when we increase the number of nodes for different network bandwidths ($B_n$). For slower networks, the throughput significantly drops when $V$ goes from 1 to 2. However, for $V\ge 2$, the throughput of shuffle consistently increases with the number of machines. Moreover, the throughput grows proportionally with the network bandwidth $B_n$. Although theoretically, the shuffle becomes more efficient with more nodes, the cost of partitioning the data and transmitting more network package headers could increase with more nodes.

\subsection{Shuffle vs. Broadcast Competitiveness}
\label{sec:models-joins}

To join the table $R$ and table $S$, we can either broadcast $R$ (assuming $R$ is the smaller table) or shuffle both $R$ and $S$. We represent the size of a table as $|\cdot|$. The time $T_b$ for broadcasting $R$ and the time $T_s$ for shuffling both are then:

\vspace*{-1ex}
    $$T_b = (N-1)\frac{|R|}{NB_n}, T_s = \frac{V-1}{V^2B_n}(|R|+|S|).$$
\vspace*{-1ex}

For $T_b < T_s$, we need:
\vspace*{-1ex}
\begin{equation}
    \label{eq:compare}
    \frac{|S|}{|R|} > \frac{N-1}{N-k}\cdot V-1.
\end{equation}

For $V=1$, the condition for the broadcast-based join to outperform the shuffle-based join is $|S|/|R| > N-1$.
If we assume a fixed $k$, then more GPUs make shuffle more favorable. On the other hand, when $V$ is small and $|S|\gg|R|$, broadcast is preferred.

\subsection{Modeling Skew}
\label{sec:model-skew}
Data skew is omnipresent in database workloads, affecting not only local processing per GPU but also data exchange. We model how data skew could affect the broadcast and shuffle operations.

\subsubsection{Broadcast}
\label{sec:broacast-model-skew}
As mentioned in Section~\ref{sec:models-broadcast}, NCCL forms rings to unify the bandwidth profiles of different links as much as possible. In a ring, messages can be pipelined, which means a GPU can stream out the message to the next hop while it is being received. This brings the benefit that broadcast with a skewed initial data placement does not cause certain links to be idle or under-utilized. In conclusion, having data skew will \emph{not} affect the broadcast as long as the broadcast is bandwidth-bound.

\subsubsection{Shuffle}
\label{sec:shuffle-model-skew}
To model the performance of shuffle under skew, we need to consider the effect of the PXN optimization~\cite{nccl-pxn}. PXN allows a GPU to use NICs of its local peers to send messages to the network. The implication of this optimization is that the data skew is visible only on a \emph{per-node} level instead of a per-GPU level. The total data sent and received through the network on any node $i$ are
\begin{equation*}
\begin{split}
    S_i &= \sum_{j\in [0,k)}\sum_{p\in [0,V)\backslash\{i\}}\sum_{q\in [0,k)} m_{ij\rightarrow pq} \text{\ (Send)} \\
    R_i &= \sum_{j\in [0,k)}\sum_{p\in [0,V)\backslash\{i\}}\sum_{q\in [0,k)} m_{pq\rightarrow ij} \text{\ (Receive)}. 
\end{split}
\end{equation*}

The time of the shuffle operation is therefore
\begin{equation*}
    T_{\text{shuffle}} = \max (S_0,...,S_{V-1}, R_0, ..., R_{V-1})/B_n.
\end{equation*}
Note that the $\text{Thpt}_{\text{shuffle}}=S/T_{\text{shuffle}}$ here is a generalization of Eq.~\ref{eq:shuffle-no-skew}.
In contrast to broadcast, data skew does have an impact on the shuffle. The performance of shuffle is determined by the \emph{node} (not the GPU) that sends or receives the largest amount of data.

\subsection{Small Message Sizes}
\label{sec:model-small-msg}
So far, we have been assuming that the efficiency of transferring a message (i.e., $B_n$, $B_g$) is independent of the message size $m$, and therefore the throughput of the shuffle and broadcast is also agnostic to message sizes. However, we observe in experiments that the message size does have an impact on the performance of both. To address this, we extend our model by parameterizing $B_n$ and $B_g$ with the message size $m$. We follow Hockney's model~\cite{wickramasinghe2016surveymethodscollectivecommunication} and assume the time to send a message via a link as $t=L+c\cdot m$, where $c$ corresponds to the time gap between sending each byte and $L$ is the latency. Therefore, we can represent $B_n$ and $B_g$ as
\vspace*{-1ex}
\begin{equation*}
    B_n(m) = \frac{m}{L_n + c_n\cdot m},  B_g(m) = \frac{m}{L_g + c_g\cdot m}.
\end{equation*}

To find the model parameters $c_n,c_g,L_n,L_g$, we can fit our analytical models against our experimental measurements. Due to the completely different implementations, shuffle and broadcast are fitted separately using the results from $V=2$.

\section{Testbed and Workloads}
\label{sec:experimental-setup}

\subsection{Cluster Configuration} 
\label{sec:cluster-config}
For this study, we use three clusters of multi-GPU Azure Virtual Machines (VMs). These clusters have GPUs from two vendors (NVIDIA, AMD) and use different interconnect technologies (NVLink, Infinity Fabric, Ethernet, InfiniBand) for communications between GPUs. Table~\ref{tab:cluster_config} lists the cluster configurations. Each VM has dual-socket CPUs with a total of 96 physical cores and $k=8$ GPUs.

\begin{table*}[ht]
\caption{Cluster Configurations. Eth: Ethernet. IB: InfiniBand.}\vspace{-8pt}
\label{tab:cluster_config}
\resizebox{\textwidth}{!}{%
\begin{threeparttable}
\begin{tabular}{|c|c|c|c|c|cc|c|c|c|c|}
\hline
\multirow{2}{*}{\textbf{Cluster}} &
  \multirow{2}{*}{\textbf{GPU Type}} &
  \textbf{HBM} &
  \multirow{2}{*}{\textbf{k}} &
  \multirow{2}{*}{\textbf{V}} &
  \multicolumn{2}{c|}{\textbf{GPU Interconnect}} &
  \multirow{2}{*}{\textbf{CPU type}} &
  \textbf{CPU} &
  \textbf{CPU} &
  \textbf{Price/hour} \\ \cline{6-7}
 &
   &
  (GiB) &
   &
   &
  \multicolumn{1}{c|}{\textbf{Intra-VM}} &
  \textbf{Inter-VM} &
   &
  \textbf{Cores} &
  \textbf{Mem (GiB)} &
  \textbf{(USD)} \\ \hline
\multirow{2}{*}{1} &
  \multirow{2}{*}{NVIDIA A100} &
  \multirow{2}{*}{80} &
  \multirow{6}{*}{8} &
  \multirow{2}{*}{7} &
  \multicolumn{1}{c|}{NVLink} &
  \multirow{2}{*}{Eth: 1x50 Gbits/sec} &
  AMD EPYC &
  \multirow{6}{*}{96} &
  \multirow{2}{*}{1800} &
  \multirow{2}{*}{32.77\tnote{*}} \\
 &
   &
   &
   &
   &
  \multicolumn{1}{c|}{300 GB/sec} &
   &
  7V12 &
   &
   &
   \\ \cline{1-3} \cline{5-8} \cline{10-11} 
\multirow{2}{*}{2} &
  \multirow{2}{*}{NVIDIA H100} &
  \multirow{2}{*}{79.6} &
   &
  \multirow{2}{*}{5} &
  \multicolumn{1}{c|}{NVLink} &
  Eth: 1x100 Gbits/sec &
  Intel Xeon &
   &
  \multirow{2}{*}{1900} &
  \multirow{2}{*}{98.32} \\ \cline{7-7}
 &
   &
   &
   &
   &
  \multicolumn{1}{c|}{450 GB/sec} &
  IB: 8x400 Gbits/sec &
  Platinum 8480C &
   &
   &
   \\ \cline{1-3} \cline{5-8} \cline{10-11} 
\multirow{2}{*}{3} &
  \multirow{2}{*}{AMD MI300X} &
  \multirow{2}{*}{191.5} &
   &
  \multirow{2}{*}{4} &
  \multicolumn{1}{c|}{Infinity Fabric} &
  Eth: 1x100 Gbits/sec &
  Intel Xeon &
   &
  \multirow{2}{*}{1850} &
  \multirow{2}{*}{63.6} \\ \cline{7-7}
 &
   &
   &
   &
   &
  \multicolumn{1}{c|}{448 GB/sec} &
  IB: 8x400 Gbits/sec &
  Platinum 8480C &
   &
   &
   \\ \hline
\end{tabular}%
\begin{tablenotes}
\item[*] This is the price for 8x200 Gbits/sec Infiniband. The Eth version, where we run our experiment, is not publicly listed.
\end{tablenotes}
\end{threeparttable}
}
\end{table*}

The communication and data exchange bandwidth between GPUs depends on their placement and available interconnect technology. We report unidirectional bandwidths for all cases.
\begin{itemize}[wide, labelwidth=!, labelindent=0pt, topsep=0pt]
\item \textbf{Intra-VM}: Communications between GPUs within the same VM can use high-bandwidth interconnects---NVIDIA NVLink for each VM in clusters 1 (A100) and 2 (H100), and AMD Infinity Fabric for each VM in cluster 3 (MI300X). For both the A100 and H100 clusters, the interconnect fabric is created with NVLinks and NVSwitches. In each VM of the A100 cluster, each GPU has 12 outgoing (and incoming) lanes, each supporting a bandwidth of up to 25 GB/sec, providing an aggregate max. outgoing (and incoming) bandwidth of 300 GB/sec. For H100, this increases to 18 outgoing (and incoming) lanes with an aggregate max. unidirectional bandwidth of 450 GB/sec. In the MI300X cluster, each GPU has 7 lanes, each providing up to 128 GB/sec bidirectional bandwidth, resulting in an aggregate max. unidirectional bandwidth of $7\times128/2 =$ 448 GB/sec.
\item \textbf{Inter-VM}: All VMs have an Ethernet NIC providing connectivity to other VMs. Additionally, every VM in the H100 and MI300X clusters has 8 Mellanox NICs, one per GPU, each supporting InfiniBand 4x NDR (Next-Generation Rate) connectivity of up to 400 Gbits/sec to other VMs. As we will show in our evaluations, inter-VM bandwidths have a major impact on scale-out performance for SQL Analytics.
\end{itemize}

In all VMs, each GPU is also connected via PCIe x16 links to the host CPU. A100 GPUs have PCIe Gen4 links, each supporting up to 31.5 GB/sec bandwidth, whereas H100 and MI300X GPUs have Gen5 links, each supporting up to 63 GB/sec. The peak CPU to main memory read bandwidths in all systems are sufficient to saturate the aggregate host-to-device PCIe bandwidth to the GPUs.

\subsection{Profiling Setup}

We report the average execution time of queries, taken over 10 runs, once each query's input data is loaded into the HBM memory of all participating GPUs. This is, measurements are taken after a warm-up phase, in which each query has been run twice to warm up the device caches.
This setup fits with our goal of demonstrating the potential speedup opportunities of multi-GPU clusters. This also represents scenarios of recurring queries, or when data loaded earlier into GPU memory may remain cached. We discuss the impact of data loading for cold runs in Section~\ref{sec:cold-runs}.

In the experiments, we break down the execution time into three parts: compute, shuffle, and broadcast. Compute time refers to the local execution time of each GPU. We measure the end-to-end query execution time, and for each shuffle/broadcast operation, we insert barriers before and after to obtain its time. We calculate the compute time by subtracting the communication time from the query execution time. 
We use vendor-specific utilities (e.g., nvidia-smi) to monitor GPU memory occupancy during query runs\footnote{We don't explicitly measure reduction operations using \texttt{ncclAllReduce} calls since these have negligible overhead.}. 

\subsection{Workloads} 
\label{sec:workloads}
We use the 22 queries of the TPC-H benchmark in our study, on both default (uniform) and skewed data. We run these queries on the TPC-H dataset, which has a mostly uniform distribution of keys, for Scale Factors (SF) of 1000 and 3000. To study the impact of data skew, we generate skewed data using the data generator for the JCC-H~\cite{jcc} dataset, but use the TPC-H queries for a close comparison with TPC-H query runs. 

We pre-partition input tables based on keys to reduce shuffles of leaf-level inputs. Since the two largest tables, \texttt{lineitem} and \texttt{orders}, are joined in queries Q3, Q4, Q5, Q7, Q8, Q9, Q10, Q12, Q18, and Q21, we partition \texttt{lineitem} by the foreign key, \texttt{l\_orderkey}, so that the distributed join can be performed locally on each GPU without first requiring a data exchange operation. For \texttt{partsupp}, we use \texttt{ps\_partkey}, and for the remaining tables, their primary keys, as the partitioning keys. Some queries, e.g., Q1 and Q6, do not need or use key-based input partitioning. 

\subsection{TQP Setup}
\label{sec:tqp-setup}

Selection of the optimal exchange operation requires knowledge of table cardinalities (Section~\ref{sec:models-joins}), as well as how data is pre-partitioned (Section~\ref{sec:workloads}). Accurate cardinality estimation of intermediate results is, however, a well-known difficult problem to solve~\cite{Abo25-cardest} and query optimizers rely on estimates which may turn out to be incorrect at run time, resulting in sub-optimal query execution or requiring corrective action at run time~\cite{Leis15-QO,Lee23-cardest}. Since in this work, we aim at showing the speedup potential with multi-GPU acceleration, once distributed TQP generates the tensor programs, we further optimize them manually (e.g., changing the data exchange operation, or the join ordering). We leave the integration with a statistic-aware distributed query optimizer
as part of our future work.

Table~\ref{tab:query-stats} shows the total number of shuffles and broadcasts in our ``optimized'' query plans for the 22 TPC-H queries.
The data exchanges involved both leaf tables (query inputs) and intermediate results. In these counts, we exclude the final gather operation for collecting partial results from the individual GPUs. Having partitioned input tables helps to reduce the number of shuffles. A different partitioning scheme would lead to a different number of data exchange operations in the resulting query plans: we discuss an example in Section~\ref{sec:non-partitioned-impact}.

\begin{table}[ht]
    \begin{flushleft}
    \caption{Exchange statistics for TPC-H query plans used.}\vspace{-8pt}
    \label{tab:query-stats}
    \begin{tabular}{@{}m{5pt}@{}|c|@{\hskip4.5pt}c@{\hskip4pt}|@{\hskip5pt}c@{\hskip4pt}|@{\hskip5pt}c@{\hskip4pt}|@{\hskip5pt}c@{\hskip4pt}|@{\hskip5pt}c@{\hskip4pt}|@{\hskip5pt}c@{\hskip4pt}|@{\hskip5pt}c@{\hskip4pt}|@{\hskip4pt}c@{\hskip4pt}|@{\hskip4pt}c@{\hskip4pt}|@{\hskip4pt}c@{\hskip4pt}|@{\hskip4pt}c@{\hskip4pt}|@{\hskip4pt}c@{\hskip4pt}|}\cline{2-14}
         &\textbf{Query}& \textbf{1}& \textbf{2}& \textbf{3}& \textbf{4}& 
                         \textbf{5}& \textbf{6}& \textbf{7}& \textbf{8}& 
                         \textbf{9}& \textbf{10}& \textbf{11}& \textbf{12}\\\cline{2-14}
         &\textbf{Shuffles}& 0& 0& 0& 0&
                            0& 0& 0& 0&
                            1& 1& 1& 0\\\cline{2-14}
         &\textbf{Broadcasts}& 0& 1& 1& 0&
                            2& 0& 2& 3&
                            2& 0& 1& 0\\\cline{2-14}                            
         
    \end{tabular}
    \begin{tabular}{@{}m{5pt}@{}|c|@{\hskip4pt}c@{\hskip4pt}|@{\hskip4pt}c@{\hskip4pt}|@{\hskip4pt}c@{\hskip4pt}|@{\hskip4pt}c@{\hskip4pt}|@{\hskip4pt}c@{\hskip4pt}|@{\hskip4pt}c@{\hskip4pt}|@{\hskip4pt}c@{\hskip4pt}|@{\hskip4pt}c@{\hskip4pt}|@{\hskip4pt}c@{\hskip4pt}|@{\hskip4pt}c@{\hskip4pt}|}\cline{2-12}
         &\textbf{Query}& \textbf{13}& \textbf{14}& \textbf{15}& 
                         \textbf{16}& \textbf{17}& \textbf{18}& \textbf{19}& 
                         \textbf{20}& \textbf{21}& \textbf{22}\\\cline{2-12}
         &\textbf{Shuffles}& 1& 1& 1&
                            1& 1& 0& 0&
                            1& 0& 1\\\cline{2-12}
         &\textbf{Broadcasts}& 1& 0& 0&
                            1& 1& 1& 1&
                            1& 1& 0\\\cline{2-12}                            
         
    \end{tabular}
    \end{flushleft}
\end{table}

\section {Data Exchange Performance}
\label{sec:data-exchange-analysis}

We develop microbenchmarks to measure the throughput of shuffle and broadcast on our platforms under various settings. These include varying the number of machines (thereby changing the number of participating GPUs and the topology of the distributed system), data skew, and message sizes. The goals are to determine the achievable throughput using NCCL and RCCL and to compare performance trends with our analytical models. 

\begin{figure}[t]
\centering
\subfloat[$V=1$. Shuffle.]{\includegraphics[width=0.5\columnwidth]{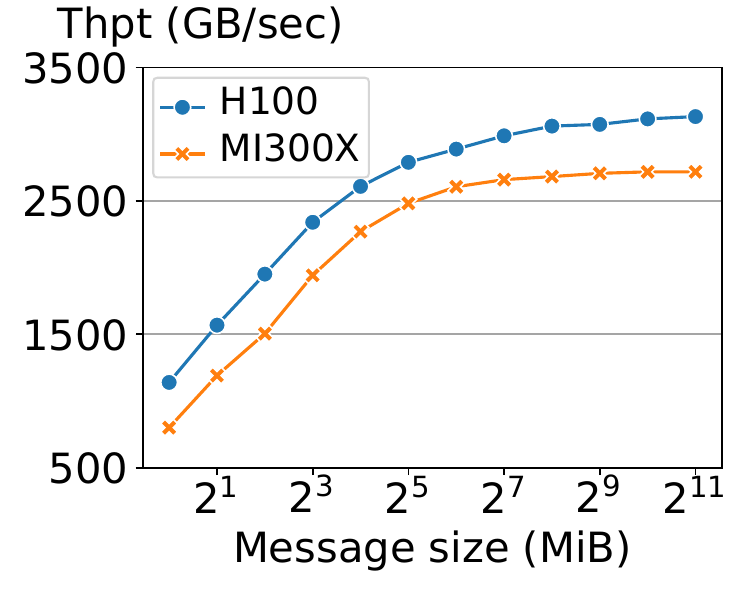}\label{fig:exch-bw-HM-V1}\vspace{-5pt}}
\subfloat[$V=4$. Shuffle.]{\includegraphics[width=0.5\columnwidth]{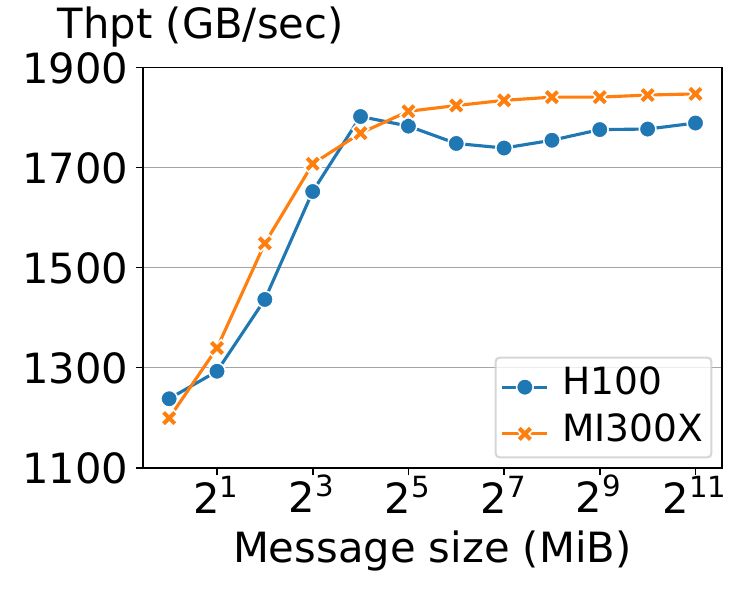}\label{fig:exch-bw-HM-V4}\vspace{-5pt}}\newline
\subfloat[$V=1$. Broadcast.]{\includegraphics[width=0.5\columnwidth]{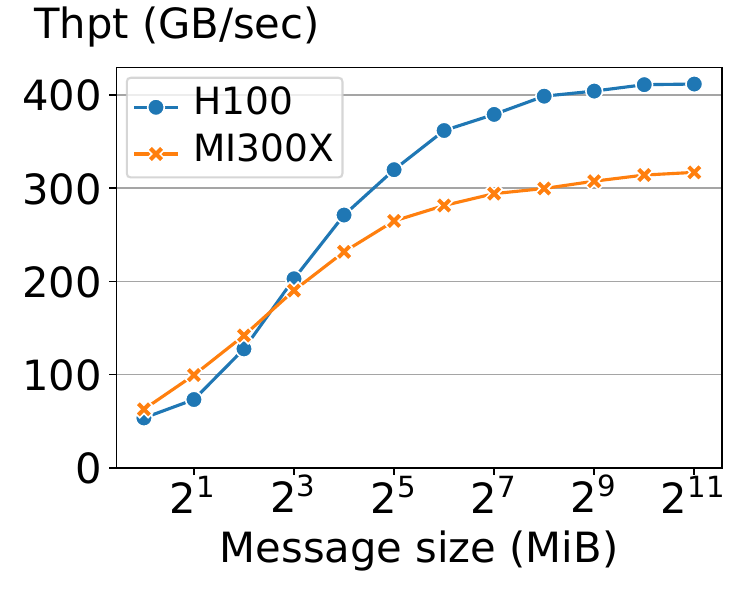}\label{fig:broadcast-bw-HM-V1}\vspace{-5pt}}
\subfloat[$V=4$. Broadcast.]{\includegraphics[width=0.5\columnwidth]{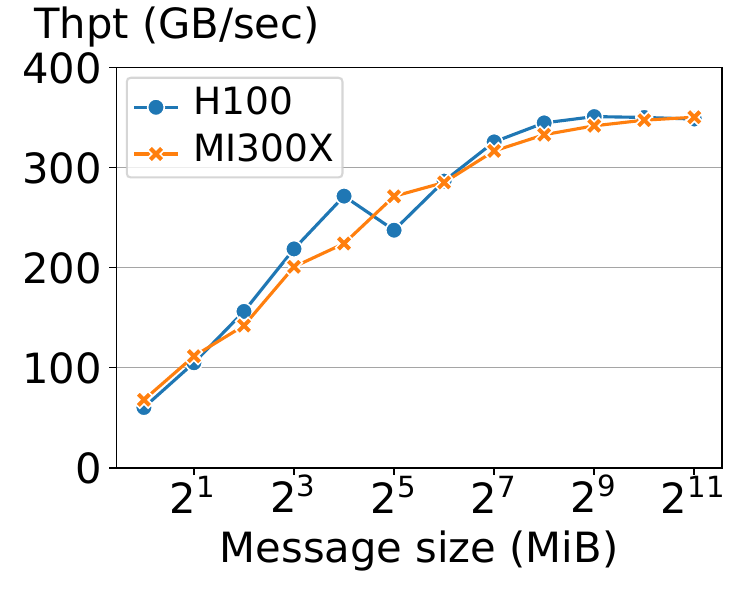}\label{fig:broadcast-bw-HM-V4}\vspace{-5pt}}
\vspace{-2ex}
\caption{Achieved throughput for shuffle and broadcast on 1 VM and 4 VMs (IB networks) of the H100 and MI300X clusters.}
\label{fig:broadcast-bw-HM}
\end{figure}

\noindent\textbf{Skew-free case.}
Figure~\ref{fig:broadcast-bw-HM} shows how the message size affects the throughput of the broadcast and shuffle without data skew.
We observe high throughputs---up to 3.1 TB/sec ($V=1$) and 1.8 TB/sec ($V=4$ with IB networks) for shuffle, and up to 412 GB/sec and 350 GB/sec for broadcast respectively.
For all cases, though, a message size that is large enough is needed to achieve the maximum throughput. For the H100+IB configuration, we need to set additional environment variables\footnote{Channels per network peer to 32 and minimum cooperative thread arrays to 24.} for NCCL without which we get a poorer broadcast but slightly better shuffle performance. 
Additionally, we see that the H100 curve is less smooth compared to MI300X's ones between message sizes of $2^4$ and $2^6$ for $V = 4$. We think that is because of NCCL switching to different algorithms.

\begin{figure}[h]
% \vspace{-2ex}
    \centering
    \subfloat[Broadcast (H100+Eth).]
    {\includegraphics[width=0.5\columnwidth]{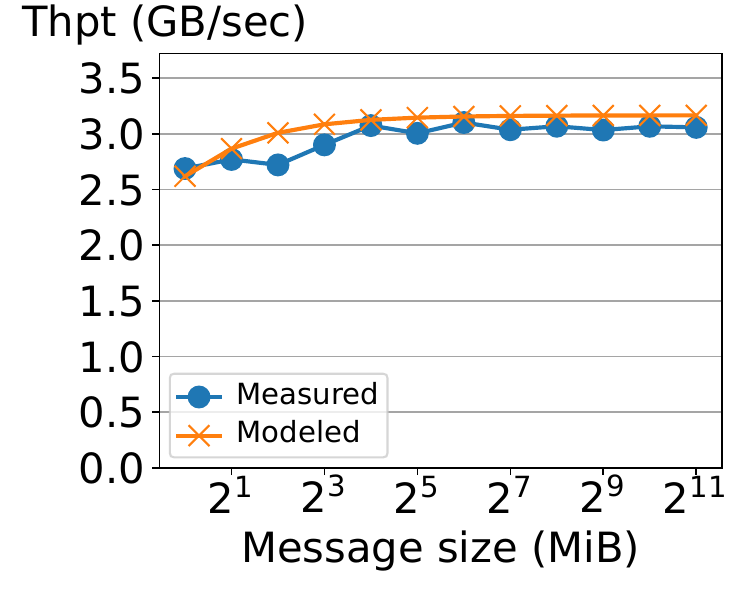}\label{fig:broadcast-model-valid-h100eth}\vspace{-4pt}}
    \subfloat[Shuffle (H100+Eth).]
    {\includegraphics[width=0.5\columnwidth]{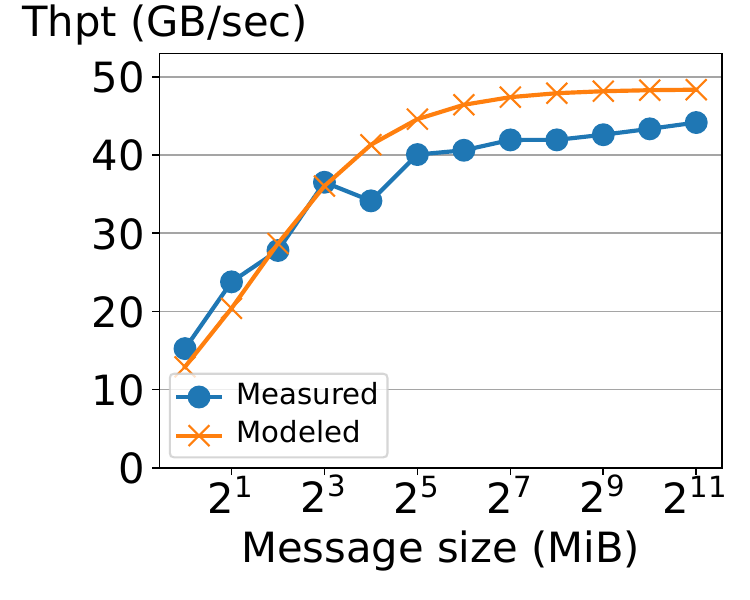}\label{fig:shuffle-model-valid-h100eth}\vspace{-4pt}}\newline
    \subfloat[Broadcast (H100+IB).]
    {\includegraphics[width=0.5\columnwidth]{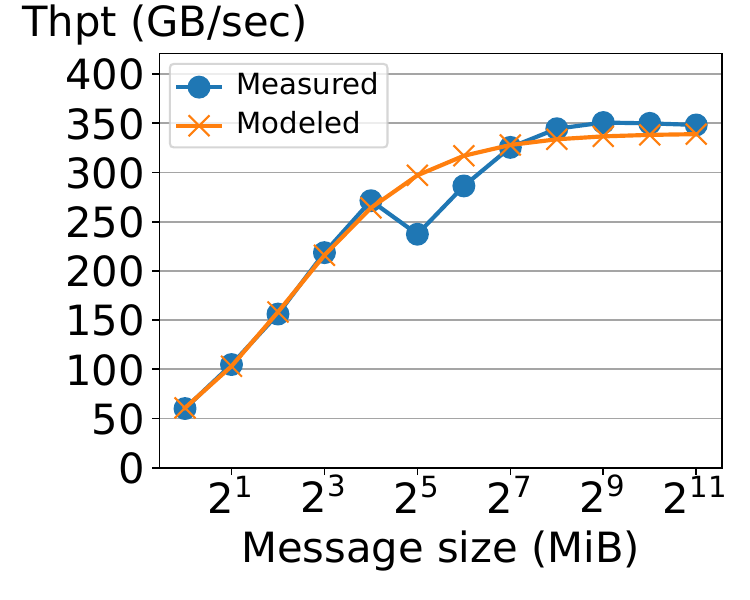}\label{fig:broadcast-model-valid-h100ib}\vspace{-4pt}}
    \subfloat[Shuffle (H100+IB).]
    {\includegraphics[width=0.5\columnwidth]{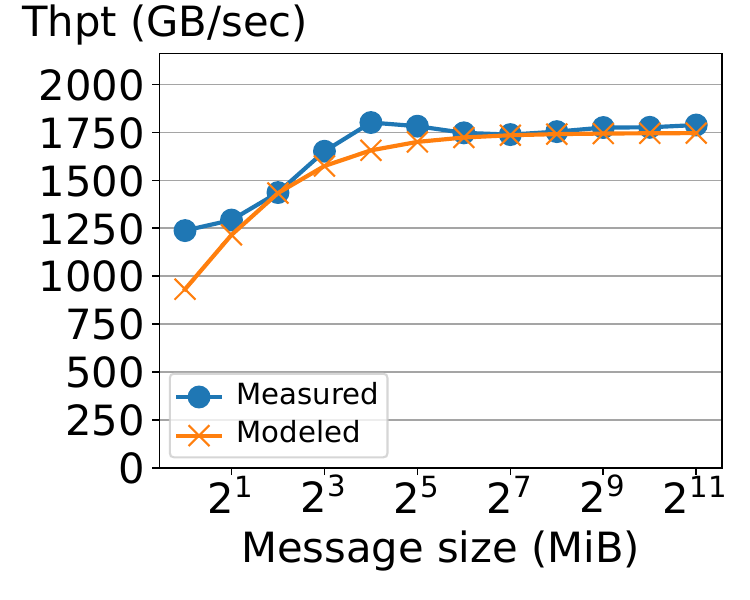}\label{fig:shuffle-model-valid-h100ib}\vspace{-4pt}}\newline
    \subfloat[Broadcast (MI300X+IB).]
    {\includegraphics[width=0.5\columnwidth]{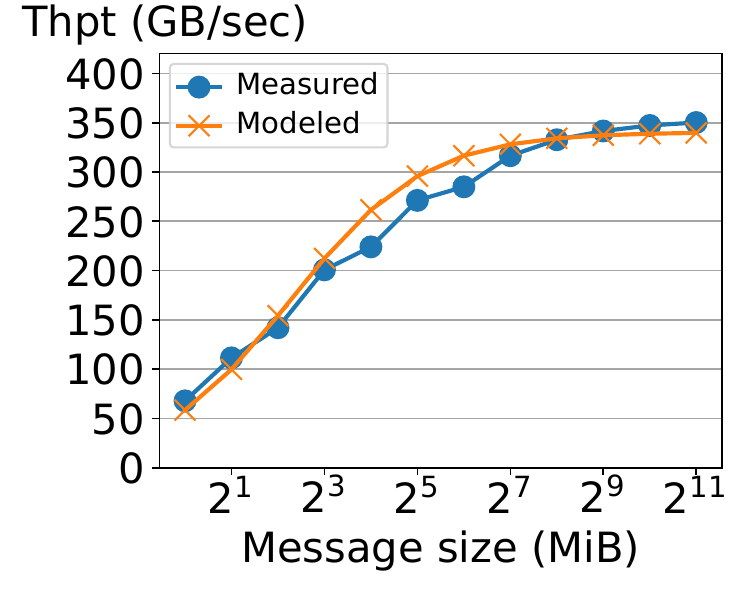}\label{fig:broadcast-model-valid-mi300ib}\vspace{-4pt}}%
    \subfloat[Shuffle (MI300X+IB).]
    {\includegraphics[width=0.5\columnwidth]{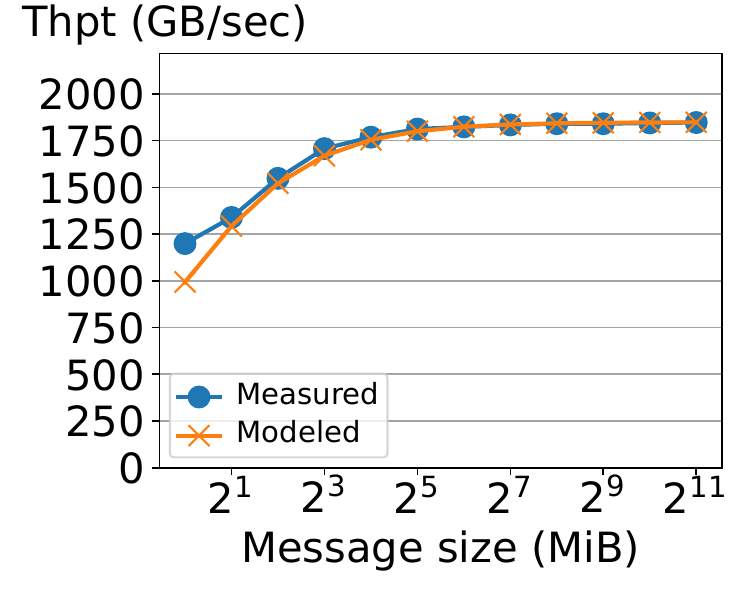}\label{fig:shuffle-model-valid-mi300ib}\vspace{-4pt}}%
    \vspace{-5pt}
    \caption{Model validation for the skew-free case. (V=4)}
    \label{fig:model-no-skew}
% \vspace{-8pt}
\end{figure}

To demonstrate the power of our model, we compare our model-predicted throughput for $V=4$ with the measured throughput in Figure~\ref{fig:model-no-skew}. To get the prediction, we first find the parameters $L_n,c_n$ (see Section~\ref{sec:message-sizes}) from $V=2$ by fitting. The results show that our models have strong predictive power for both shuffle and broadcast, for varying message sizes, for different GPUs, and for different network types. The implication of this is that we can use our model to accurately predict the data exchange performance of real query workloads given a cluster configuration. The prediction provides insights into whether the user should scale out the cluster to gain more performance. As we will see in Section~\ref{sec:perf-projection}, although scaling out brings more parallelism, it also makes data exchange less efficient due to smaller message sizes.

\noindent\textbf{Skew case.} 
To validate our models in the skew case, we introduce skew to the initial data placement. Assuming GPUs $G_0, G_1, ..., G_{N-1}$ have $x, x+fx, ..., x+(N-1)fx$ data, respectively. Here, we call $f$ the ``skew gradient''. Varying $f$ produces different levels of skewness, and when $f=0$, there is no skew. We fix the total dataset size to be $S=N\times1\text{ GiB}$; therefore, for each $f$, we need to find out $x$. For example, if $f=1,\ V=4,\ k=8$, then the first GPU will have around 62 MiB data, whereas the last GPU will have around 1.94 GiB data. In contrast, if $f=0$, every GPU will have 1 GiB of data.

\begin{figure}[t]
    \centering
    \subfloat
    {\includegraphics[width=0.49\columnwidth]{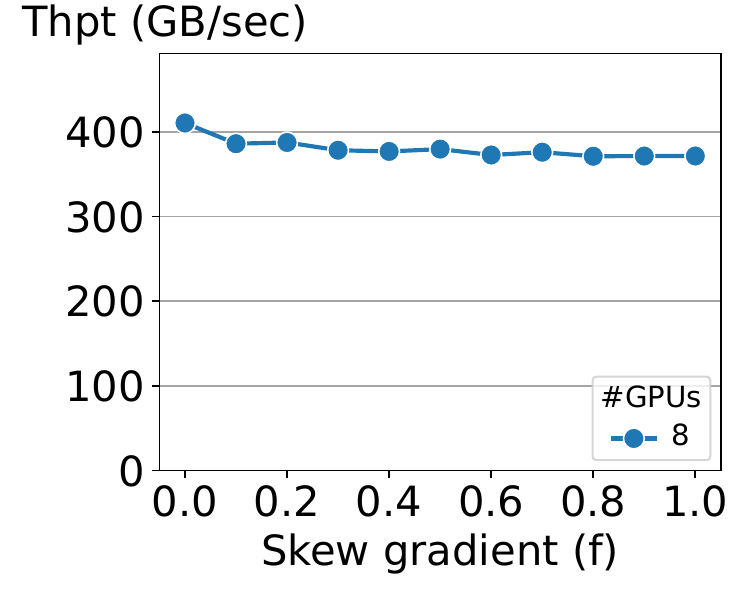}\label{fig:broadcast-model-skew-V1}\vspace{-4pt}}
    \subfloat
    {\includegraphics[width=0.49\columnwidth]{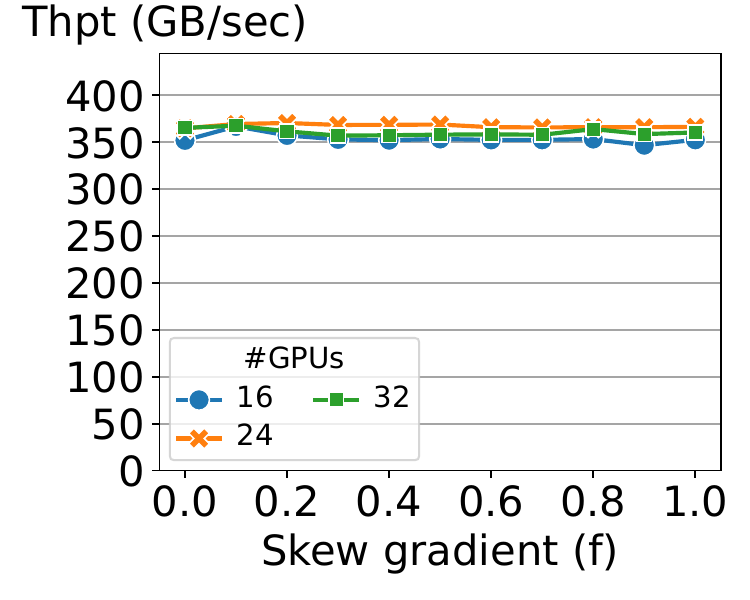}\label{fig:broadcast-model-skew-V234}\vspace{-4pt}}
    \vspace{-8pt}
    \caption{Broadcast + data skew.}
    \label{fig:broadcast-model-skew}
% \vspace{-10pt}
\end{figure}

We measure the performance of broadcast for different $V$ and $f$. The result (Figure~\ref{fig:broadcast-model-skew}) shows that the throughput of broadcast is not affected by the data skew, which agrees with our model.

\begin{figure}[t]
    \centering
    \subfloat[Skew across VMs (V=2)]
    {\includegraphics[width=0.49\columnwidth]{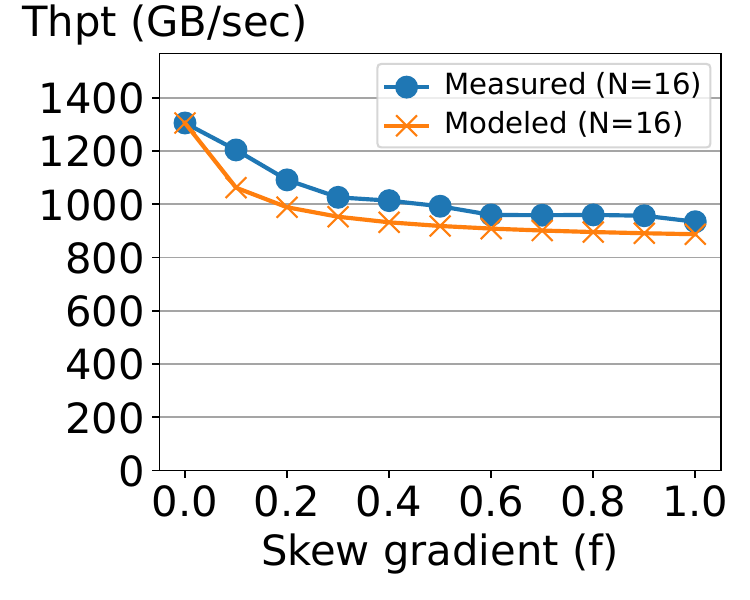}\label{fig:shuffle-model-skew-V2}\vspace{-8pt}}
    \subfloat[Skew across VMs (V=3)]
    {\includegraphics[width=0.49\columnwidth]{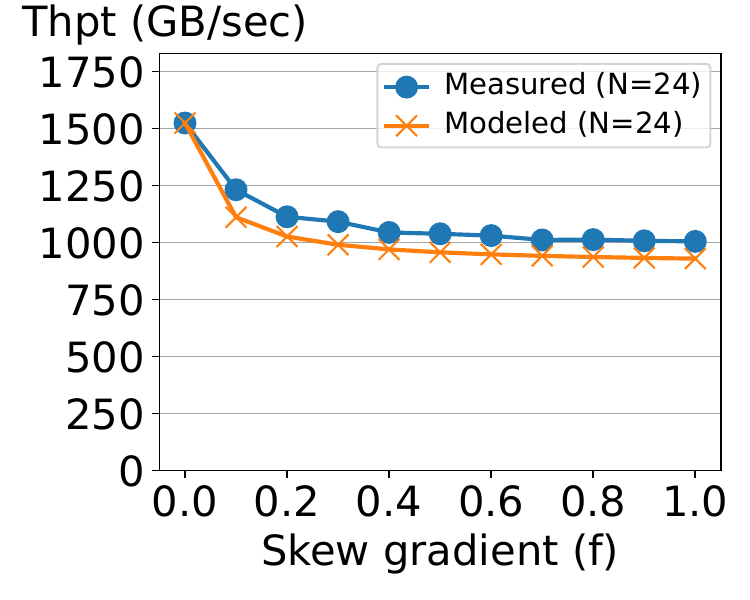}\label{fig:shuffle-model-skew-V3}\vspace{-8pt}}\newline
    \subfloat[Skew across VMs (V=4)]
    {\includegraphics[width=0.49\columnwidth]{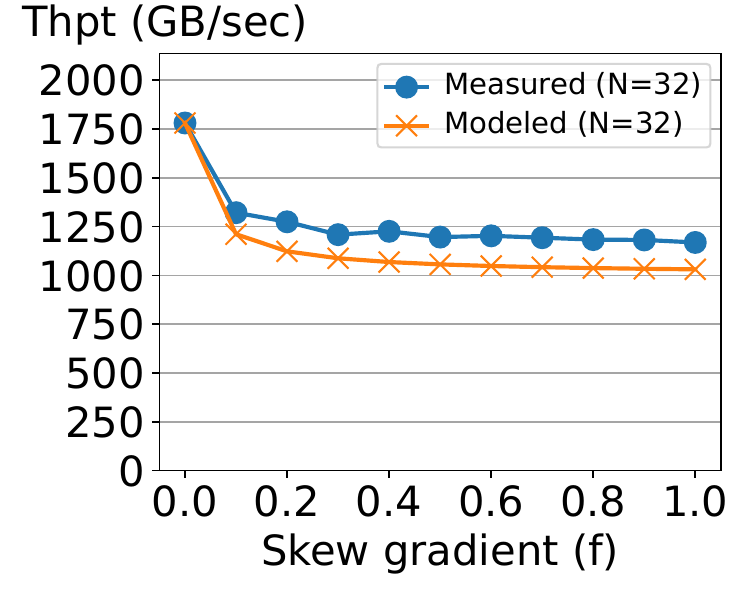}\label{fig:shuffle-model-skew-V4}\vspace{-8pt}}
    \subfloat[Skew within VM only (V=4).]
    {\includegraphics[width=0.49\columnwidth]{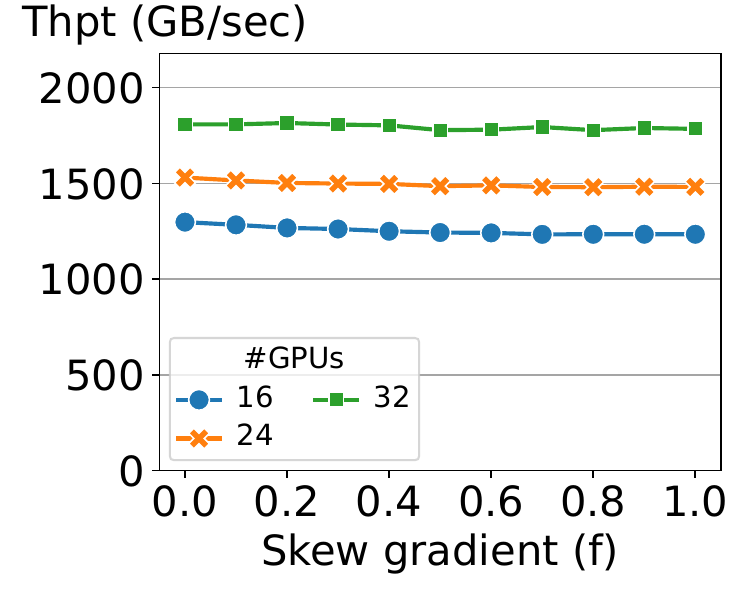}\label{fig:shuffle-partial-skew-model}\vspace{-8pt}}
    \vspace{-8pt}
    \caption{Shuffle + data skew.}
    \label{fig:shuffle-model-skew}
    \vspace{-3ex}
\end{figure}

In the shuffle case, on top of the initial data placement, we let each GPU send the same amount of data to all of its receivers. In this case, the last node sends the largest amount of data, and the first node receives the largest amount of data. Figure~\ref{fig:shuffle-model-skew-V2}-\ref{fig:shuffle-model-skew-V4} shows our modeled shuffle throughput in this case in comparison with the actually measured throughput. We find the $B_n$ parameter from fitting the $f=0$ data point. The result shows that our model also captures the effect of skew very well. The result also implies that the performance of shuffle degrades noticeably with the existence of skew across VMs. According to our model, if the skew only exists across GPUs but not VMs, the skew does not have an effect. We further validate this claim by letting each GPU in each VM contain $x, x+fx, ..., x+7fx$ data so that there is only skew intra-VM but not inter-VM. Results in Figure~\ref{fig:shuffle-partial-skew-model} show that the skew in this case does not affect the shuffle performance.

\section{TPC-H Performance Analysis}
\label{sec:tpch-analysis}

In this section, we discuss the times for TPC-H queries on 1 TB and 3 TB datasets with distributed TQP. We analyze the results from various angles, including run times (Section 6.1), time breakdown  (Section 6.2), price performance  (Section 6.4), data exchange (Section 6.5), and memory utilization  (Section 6.6). Furthermore, based on existing results, we project the performance for future networks and scale out with more VMs (Section 6.3). The goal is to characterize the execution of a classic SQL workload and offer insights into the efficiency and bottlenecks of multi-GPU clusters. 

\subsection{Workload performance}
\label{sec:tpch-overall-performance}
\begin{table*}[t]
\caption{Query Completion Summary for TPC-H.}
\label{tab:query-completion-compressed}
\vspace{-8pt}
\resizebox{\textwidth}{!}{%
\begin{tabular}{@{}lllllllllllllllllllllllll@{}}
\toprule
Cluster & Benchmark & $V$ & Q1 & Q2 & Q3 & Q4 & Q5 & Q6 & Q7 & Q8 & Q9 & Q10 & Q11 & Q12 & Q13 & Q14 & Q15 & Q16 & Q17 & Q18 & Q19 & Q20 & Q21 & Q22 \\ \midrule
All & TPCH 1 TB & $\ge 1$ & \cmark & \cmark & \cmark & \cmark & \cmark & \cmark & \cmark & \cmark & \cmark & \cmark & \cmark & \cmark & \cmark & \cmark & \cmark & \cmark & \cmark & \cmark & \cmark & \cmark & \cmark & \cmark \\ \midrule
All & TPCH 3 TB & 1 & \textcolor{red}{\xmark} & \cmark & \textcolor{red}{\xmark} & \textcolor{red}{\xmark} & \textcolor{red}{\xmark} & \cmark & \textcolor{red}{\xmark} & \textcolor{red}{\xmark} & \textcolor{red}{\xmark} & \textcolor{red}{\xmark} & \cmark & \textcolor{red}{\xmark} & \cmark & \textcolor{red}{\xmark} & \textcolor{red}{\xmark} & \cmark & \textcolor{red}{\xmark} & \textcolor{red}{\xmark} & \textcolor{red}{\xmark} & \textcolor{red}{\xmark} & \textcolor{red}{\xmark} & \cmark \\ 
A100/H100 & TPCH 3 TB & 2 & \textcolor{red}{\xmark} & \cmark & \textcolor{red}{\xmark} & \cmark & \textcolor{red}{\xmark} & \cmark & \textcolor{red}{\xmark} & \textcolor{red}{\xmark} & \textcolor{red}{\xmark} & \cmark & \cmark & \cmark & \cmark & \cmark & \cmark & \cmark & \cmark & \cmark & \cmark & \cmark & \textcolor{red}{\xmark} & \cmark \\ 
A100/H100 & TPCH 3 TB & 3 & \cmark & \cmark & \cmark & \cmark & \cmark & \cmark & \textcolor{red}{\xmark} & \textcolor{red}{\xmark} & \cmark & \cmark & \cmark & \cmark & \cmark & \cmark & \cmark & \cmark & \cmark & \cmark & \cmark & \cmark & \cmark & \cmark \\ 
A100/H100 & TPCH 3 TB & $\ge 4$ & \cmark & \cmark & \cmark & \cmark & \cmark & \cmark & \cmark & \cmark & \cmark & \cmark & \cmark & \cmark & \cmark & \cmark & \cmark & \cmark & \cmark & \cmark & \cmark & \cmark & \cmark & \cmark \\ \midrule
MI300X & TPCH 3 TB & $\ge 2$ & \cmark & \cmark & \cmark & \cmark & \cmark & \cmark & \cmark & \cmark & \cmark & \cmark & \cmark & \cmark & \cmark & \cmark & \cmark & \cmark & \cmark & \cmark & \cmark & \cmark & \cmark & \cmark \\ \bottomrule
\end{tabular}%
}
\vspace{1ex}
\end{table*}

\begin{figure}[t]
    \centering
    \subfloat[TPC-H SF=1000 (1TB)]{ 
        \includegraphics[trim={0ex 3ex 0ex 0ex},clip,width=.95\columnwidth]{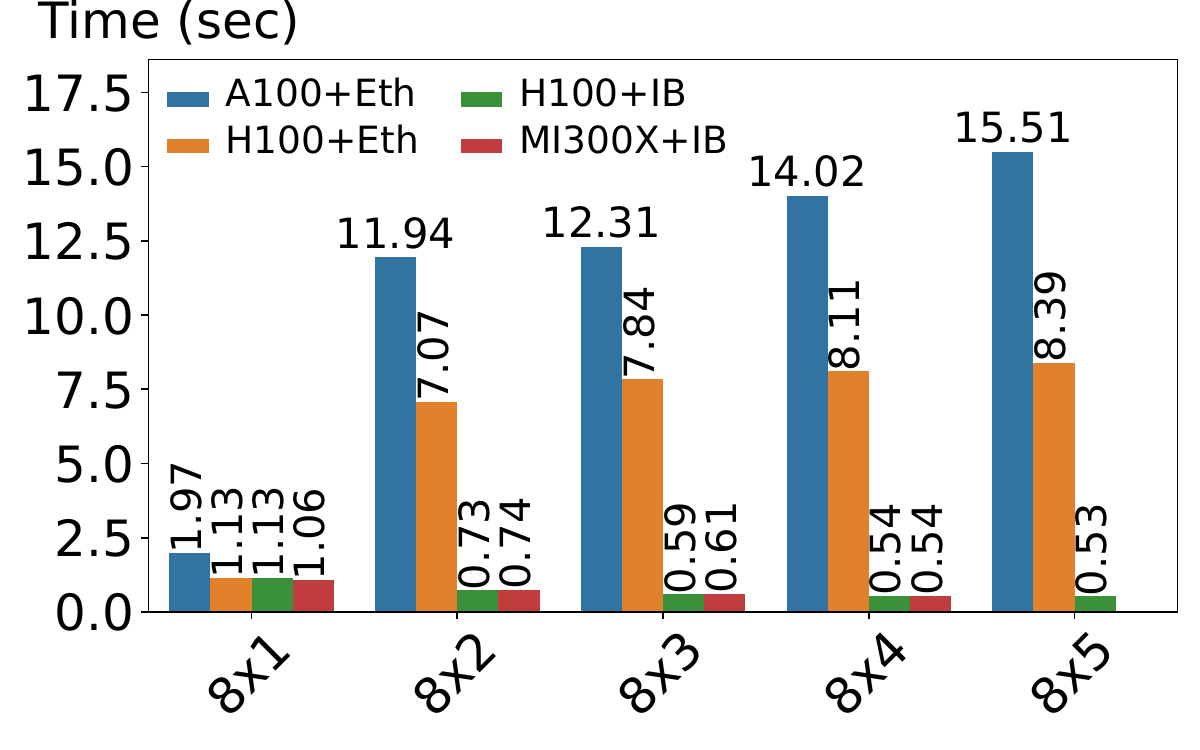}
        \label{fig:sku_compare_tpch1000}
        \vspace{-3pt}
    }\newline
    \subfloat[TPC-H SF=3000 (3TB)]{
         \includegraphics[trim={0ex 3ex 0ex 0ex},clip,width=.95\columnwidth]{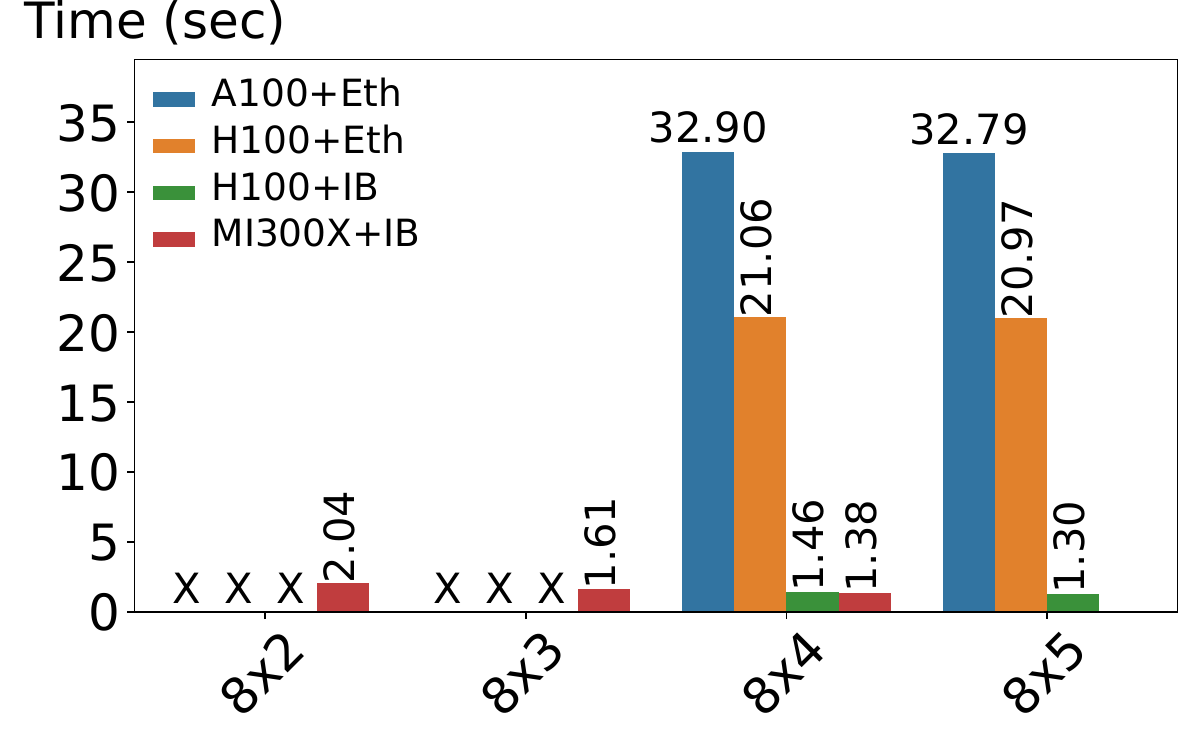}
         \label{fig:sku_compare_tpch3000}
         \vspace{-3pt}
    }\vspace{-5pt}
    \caption{Workload performance (sum of run times for 22 queries) with different cluster configurations.\vspace{-5pt}}
    \label{fig:sku_compare}
    \vspace{-10pt}
\end{figure}

\begin{figure*}[t]
    \centering
    \subfloat[A100+Ethernet]{
        \includegraphics[trim={0ex 4ex 0ex 0ex},clip,width=0.33\textwidth]{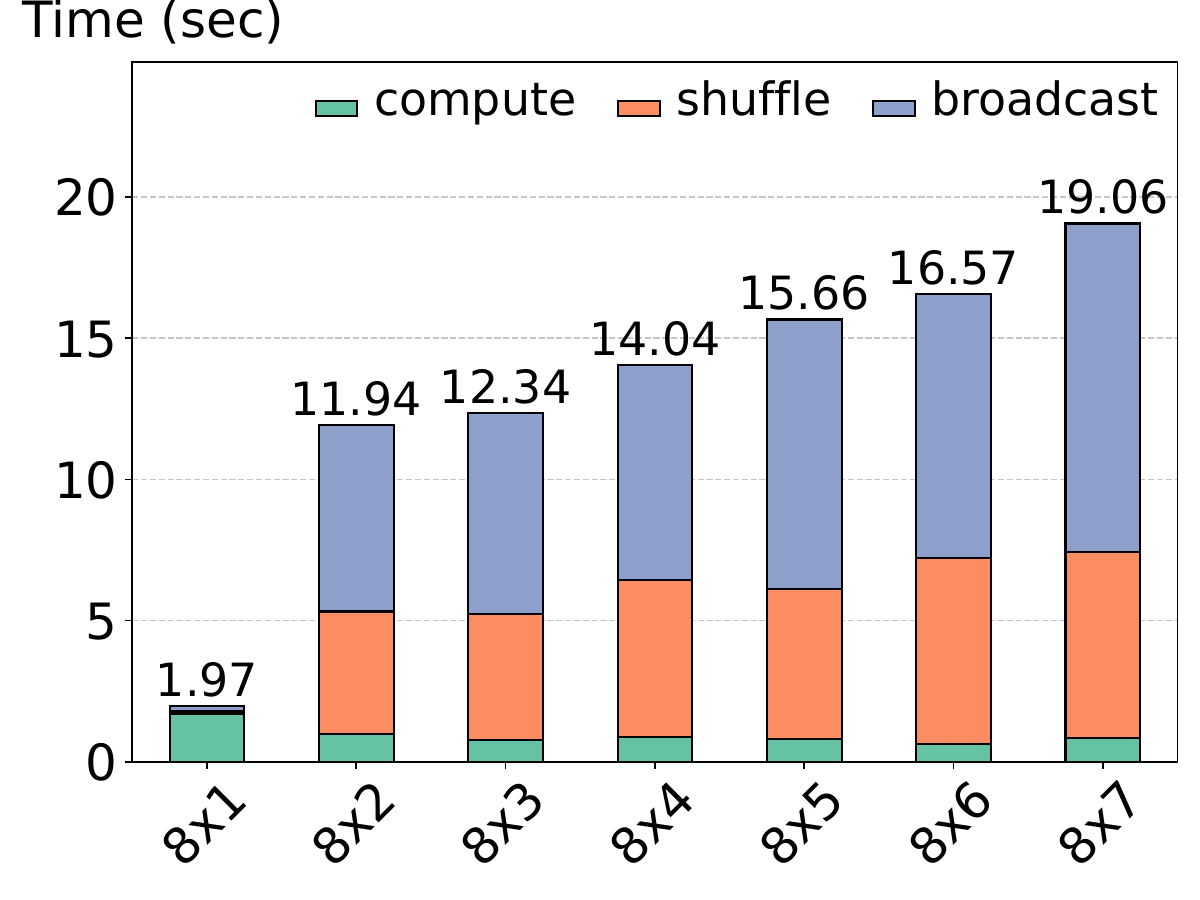}
         \label{fig:total_time_breakdown_tpch1000_a100_eth}
    }
    \subfloat[H100+Ethernet]{
         \includegraphics[trim={0ex 4ex 0ex 0ex},clip,width=0.33\textwidth]{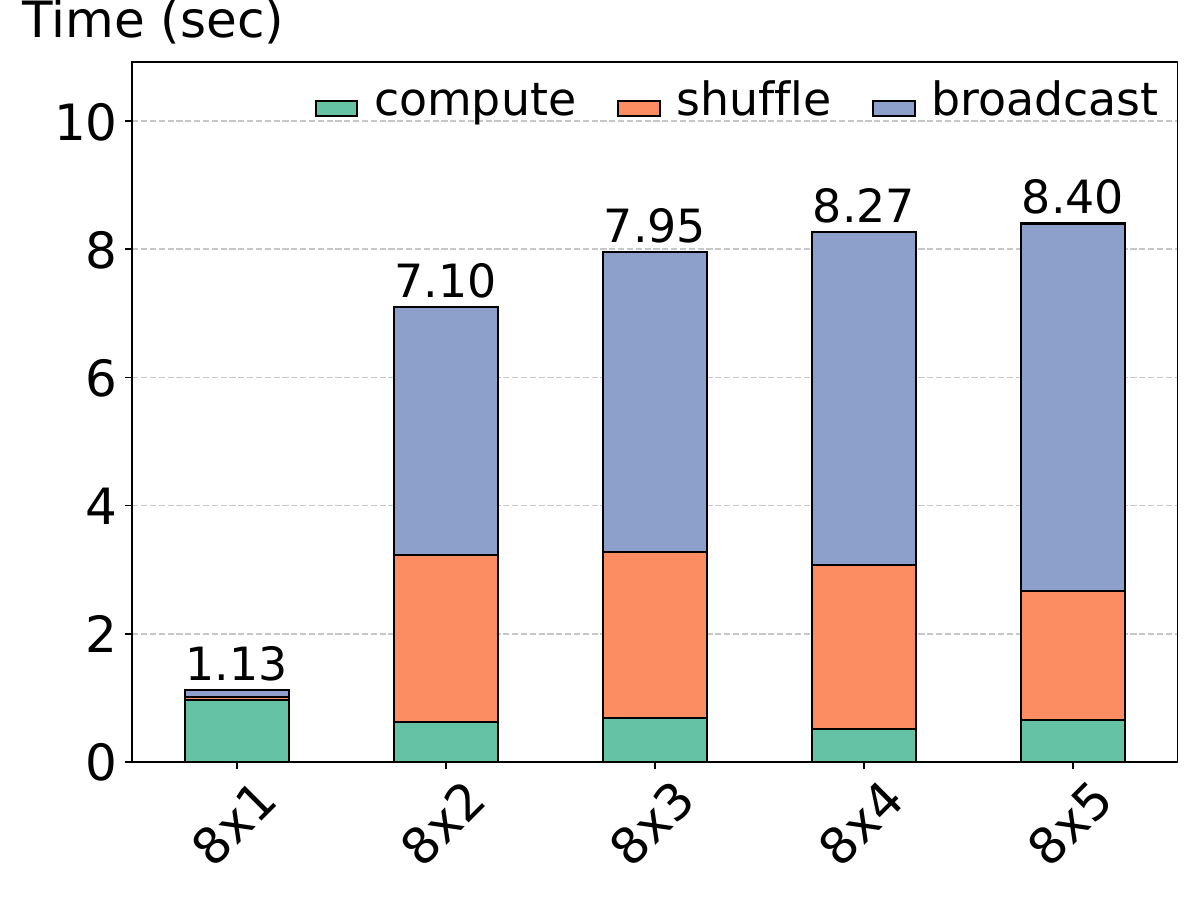}
         \label{fig:total_time_breakdown_tpch1000_h100_eth}
    }
    \subfloat[H100+InfiniBand]{
         \includegraphics[trim={0ex 4ex 0ex 0ex},clip,width=0.33\textwidth]{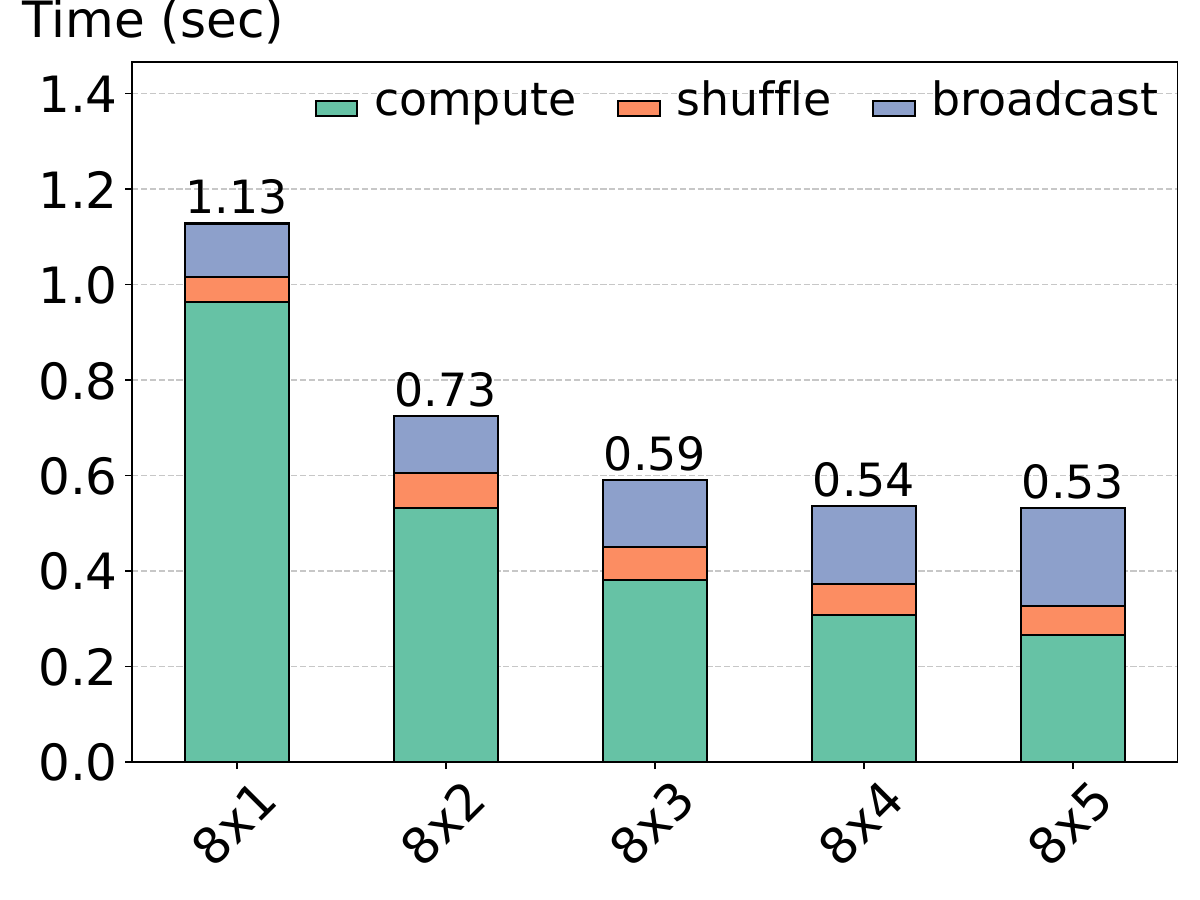}
         \label{fig:total_time_breakdown_tpch1000_h100_ib}
    }
    \vspace{-2ex}
    \caption{22-query Total Time Breakdown. TPC-H, SF=1000. (Each bar from bottom up: compute, shuffle, and broadcast.)}
    \label{fig:total_time_breakdown_tpch1000}
\end{figure*}

Figure~\ref{fig:sku_compare} shows overall workload performance, which is the sum of the warm run times over all 22 TPC-H queries. We only show cluster configurations where all queries can be run. For SF=1000, this was possible on all cluster configurations, but SF=3000 required $V\ge 4$ for A100 and H100 VMs
, and $V\ge 2$ for MI300X VMs. Table~\ref{tab:query-completion-compressed} lists query completions for all configurations.

We see that all queries with SF=1000 could be run in 1.1 seconds using 8 GPUs in a single VM.
With multiple VMs and more GPUs, further speedups are possible. With 40 GPUs in 5 VMs with IB interconnects, the total run time reduces to 0.53 seconds for SF=1000 and 1.3 seconds for SF=3000, representing more than two orders of magnitude in speedup over published numbers for these scale factors using CPU-only machines~\cite{HPE_TPC-H_2022,HPE_TPC-H_3TB_2024}. Interestingly, the run times are comparable between H100 and MI300X machines, even though they are from different vendors.

The above speedups with multiple VMs are possible only with high-bandwidth networks, such as the IB networks that we use. With Ethernet connectivity, the times are significantly larger for multi-VM configurations, e.g., 15.51 seconds for A100 and 8.39 seconds for H100 at $V=5$. We omit the numbers for MI300X with Ethernet from the figure since the network configuration is similar to that of the H100 cluster, and the single-VM performances are also similar. Multi-VM configurations without high-bandwidth interconnects only increase run time and costs, and thus are not an efficient setup for these workloads.

\subsection{Time breakdown}
\label{sec:time-breakdown}
Figure~\ref{fig:total_time_breakdown_tpch1000} shows the breakdown of workload run time (sum of run times of the 22 queries) into compute, shuffle, and broadcast components. For the multi-VM InfiniBand configurations, we only show breakdowns for the H100 cluster since the network configuration and total run times for the MI300X cluster are similar.

If only a single VM is used (8x1 configuration), then the compute time dominates. 
As the number of VMs ($V$) increases, shuffle and broadcast times become major contributors to overall time. 
This is because of the relatively low inter-VM network bandwidths compared to intra-VM NVLink bandwidths---two orders of magnitude lower for Ethernet and one order of magnitude lower for InfiniBand---which severely impact the performance of data exchange operations. 
For 5 VMs (8x5 configuration), they contribute 55.1\%, 92.3\%, and 94.8\% of times for H100+IB, H100+Eth, and A100+Eth configurations. 
The significantly lower-bandwidth Ethernet NICs on the A100 VMs cause data exchanges to take the longest time on the A100 cluster. 
The time breakdowns highlight the critical importance of network bandwidth for the scale-out performance of distributed GPU-based analytical query processing. 

Between shuffle and broadcast, the latter dominates as we scale out by increasing $V$. 
For example, for the A100+Eth, shuffle and broadcast contribute 34.5\% and 61\% respectively to the overall time for $V=7$, but 36.4\% and 55.4\% for $V=2$. 
For H100+IB configurations, they contribute 11.3\% and 38.7\% for $V=5$, and 10.2\% and 16.5\% for $V=2$.
This trend is due to the poor scalability of broadcast with $V$ compared to shuffle, as we discussed in Section~\ref{sec:models}. 

\begin{figure}[t]
    \centering
    \begin{subfigure}[b]{0.45\columnwidth}
         \centering
         \includegraphics[trim={0ex 4ex 0ex 0ex},clip,width=\textwidth]{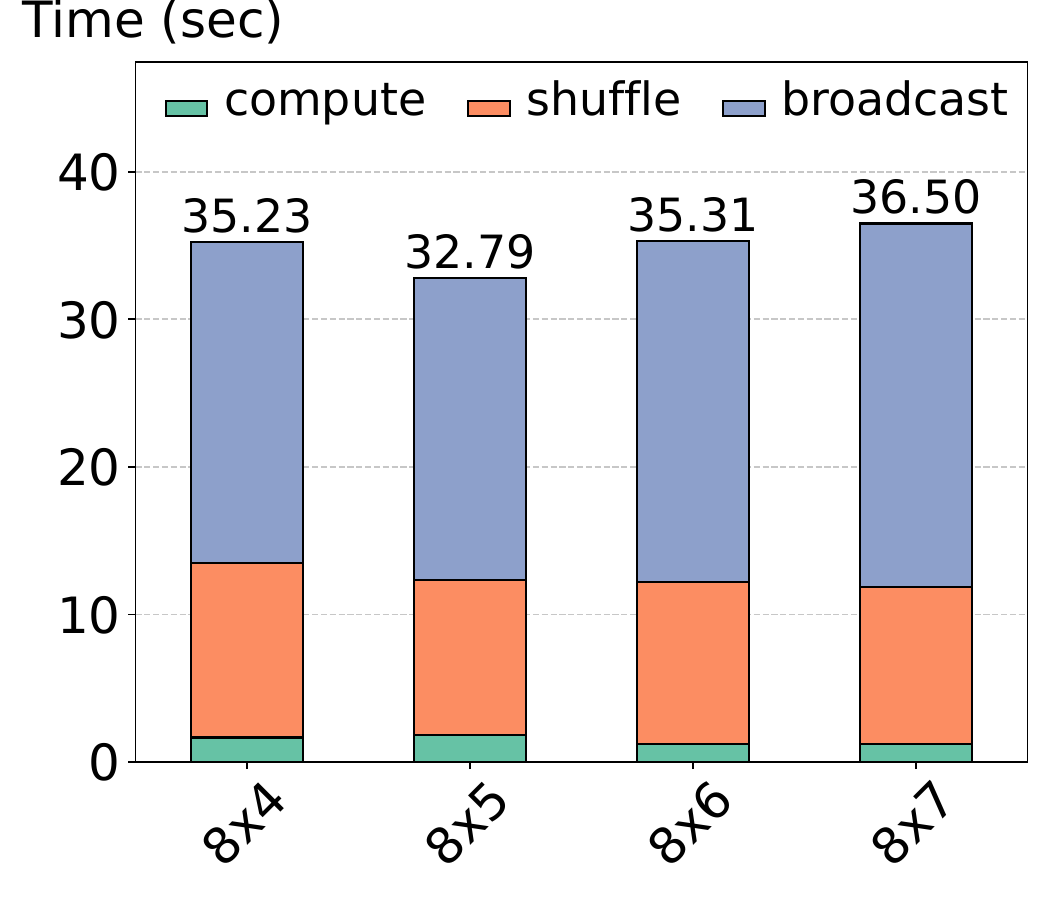}
         \caption{A100+Eth}
         \label{fig:total_time_breakdown_tpch3000_a100_eth}
    \end{subfigure}
    %\hfill
    \begin{subfigure}[b]{0.26\columnwidth}
         \centering
         \includegraphics[trim={0ex 4ex 0ex 0ex},clip,width=\textwidth]{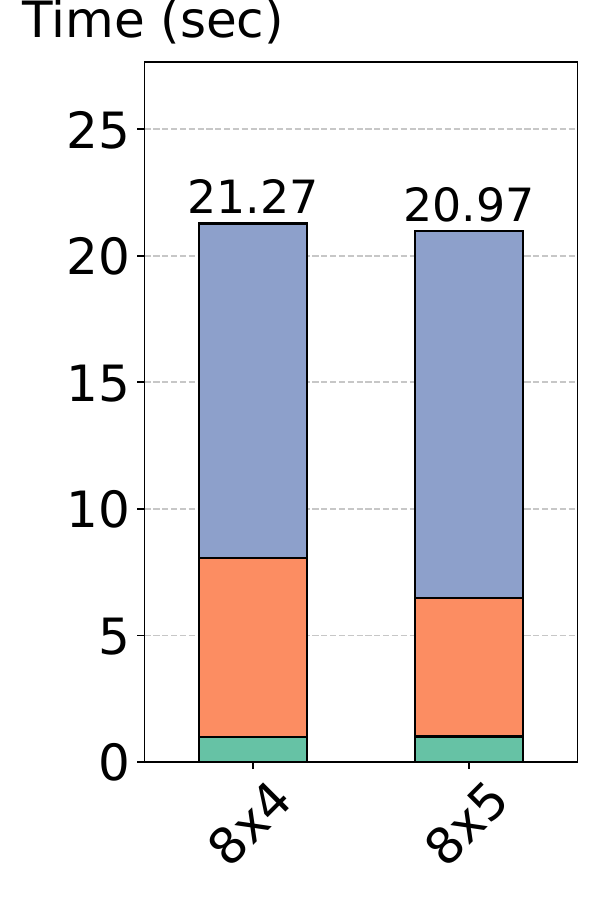}
         \caption{H100+Eth}
         \label{fig:total_time_breakdown_tpch3000_h100_eth}
    \end{subfigure}
    %\hfill
    \begin{subfigure}[b]{0.26\columnwidth}
         \centering
         \includegraphics[trim={0ex 4ex 0ex 0ex},clip,width=\textwidth]{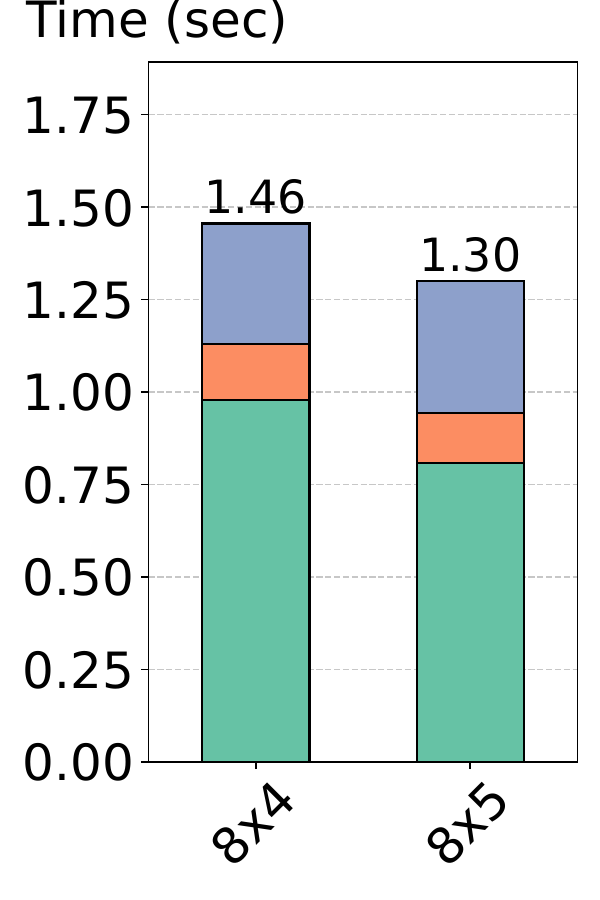}
         \caption{H100+IB}
         \label{fig:total_time_breakdown_tpch3000_h100_ib}
    \end{subfigure}
    \vspace{-2ex}
    \caption{22-query Total Time Breakdown. TPC-H, SF=3000.}
    \label{fig:total_time_breakdown_tpch3000}
\end{figure}

Figure~\ref{fig:total_time_breakdown_tpch3000} shows a breakdown of total run times for TPC-H SF=3000, on the A100 and H100 clusters for values of $V\ge 4$ where all 22 queries were completed. Since the MI300X GPUs have more HBM, we can run all 22 queries on the MI300X cluster for $V\ge 2$. Similar to SF=1000 breakdowns, these breakdowns also show the very significant impact of network performance on shuffles and broadcasts, and consequently, on overall workload performance. 

We also note the run time scales sub-linearly with the increase in scale factor. 
For the H100+IB configuration, the run time increased by 2.7$\times$ and 2.5$\times$ for $V=4$ and $V=5$ respectively, compared to the corresponding times for SF=1000.

\subsection{Performance Projection}
\label{sec:perf-projection}
We project the performance of TPC-H 1TB for an increasing number of machines ($V$), using our analytical scalability models for shuffle and broadcast. 
One of the most important goals of our modeling effort in Section~\ref{sec:models} is to help users build an expectation on how the TPC-H performance will change before they decide to scale out. 
We discuss two different methods of predicting the TPC-H performance, ``best-accuracy'' and ``best-effort''. Both utilize our analytical models in the same way, and the only difference is how they predict the compute time. The ``best-accuracy'' method predicts the compute time by fitting the five measurements of compute time from $V=1$ to $5$ into a power model, $\text{Compute}=a\times N^{b}\ (-1< b<0)$, which captures the sub-linearity. However, this requires the user to run the TPC-H on more than one machine, which is potentially costly and unavailable. To solve this, the ``best-effort'' method only uses the compute time at $V=1$ and assumes a perfect linear scaling of the compute time with respect to $V$.

\begin{figure}[t]
     \centering
     \subfloat[Broadcast (V=4)]
    {\includegraphics[width=0.49\columnwidth]{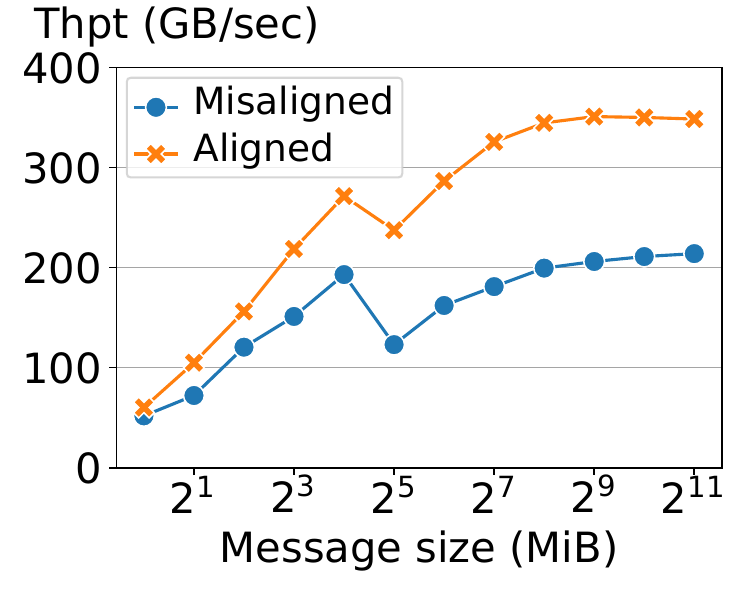}\label{fig:broadcast-misalign}\vspace{-5pt}}
    \subfloat[Shuffle (V=4)]
    {\includegraphics[width=0.49\columnwidth]{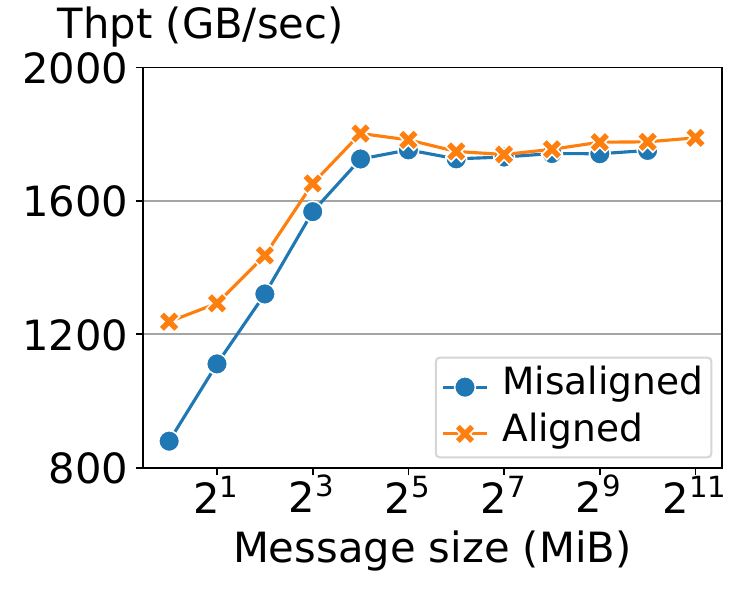}\label{fig:shuffle-misalign}\vspace{-5pt}}
     \caption{Effect of misalignment on shuffle and broadcast.\vspace{-5pt}}
     \label{fig:misalign}
\end{figure}

To project the shuffle and broadcast time in both methods, we leverage (1) the workset size of each shuffle and broadcast and (2) the $c_n,c_g,L_n,L_g$ constants obtained from the microbenchmarks ($V=1,2$), which are essential to account for small message sizes. We argue that these are reasonable and accessible inputs to our models: (1) can be obtained by running the queries for $V=1$, and (2) can be obtained from fact sheets or published literature.
\textbf{(Projection I)} This vanilla approach ignores the effect of message sizes and uses the $B_n$ and $B_g$ in Table~\ref{tab:cluster_config} to calculate the throughput of shuffle and broadcast according to Equation~\ref{eq:broadcast-no-skew} and~\ref{eq:shuffle-no-skew}. Since in real systems the theoretical peak bandwidth of a link cannot be attained, we normalize our predicted performance by making it agree with the measured performance at $V=1$. 
\textbf{(Projection II)} This approach (``+Small Msg'') takes the average message size of each shuffle and broadcast as input to $B_n(m)$ and $B_g(m)$ and then applies Equation~\ref{eq:broadcast-no-skew} and~\ref{eq:shuffle-no-skew}. 
\textbf{(Projection III)} This approach (``+Small Msg+Misalignment'') considers yet another influential factor that is often seen in real-world workloads, namely misaligned start addresses of send/receive buffers. As shown in Figure~\ref{fig:misalign}, if the starting addresses of the send or receive buffer are not aligned at 16 bytes, the performance will degenerate substantially in some cases. 

\begin{figure}[t]
     \centering
     \begin{subfigure}[b]{.66\columnwidth}
     \includegraphics[trim={0ex 4ex 0ex 0ex},clip,width=\textwidth]{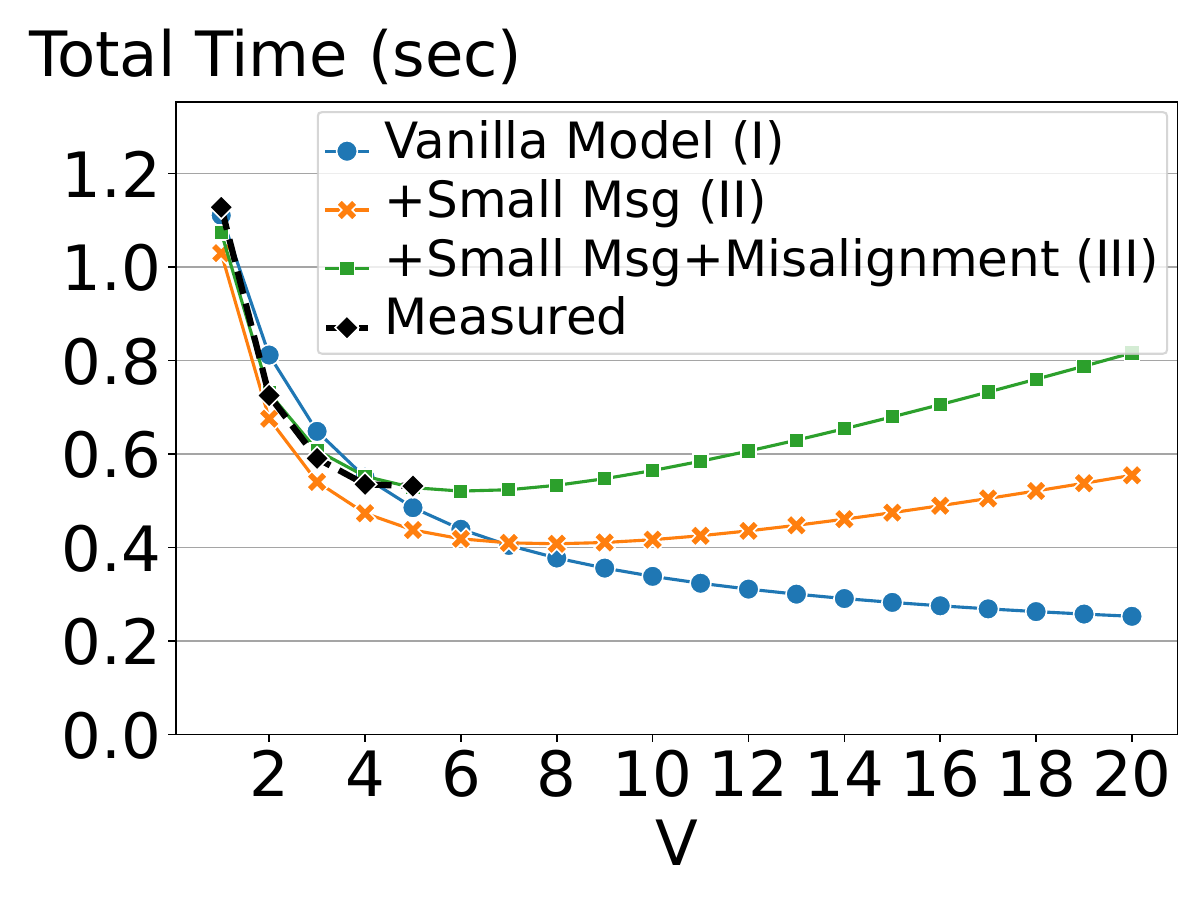}
     \vspace{-3ex}
     \caption{Models for TPC-H projections.}%
     \label{fig:tpch1000-proj-methods}
     \end{subfigure}%
     \begin{subfigure}[b]{.33\columnwidth}
     \includegraphics[trim={0ex 4ex 0ex 0ex},clip,width=\textwidth]{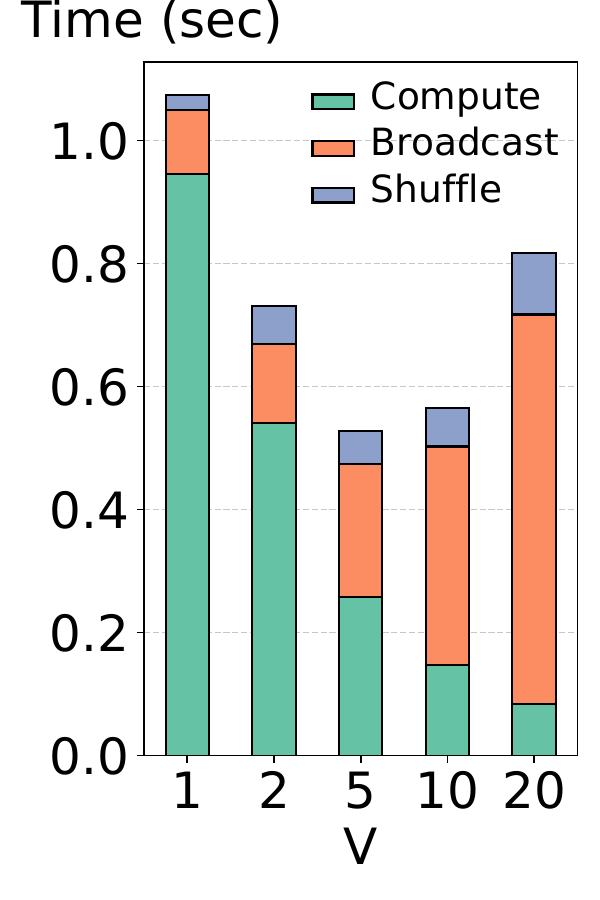}
     \vspace{-3ex}
     \caption{Project breakdown.}
     \label{fig:tpch1000-proj-breakdown}
     \end{subfigure}
     \vspace{-2ex}
     \caption{Project TPC-H performance with our models.}
     \label{fig:tpch1000-proj}
     \vspace{-1ex}
\end{figure}

Figure~\ref{fig:tpch1000-proj-methods} shows the predicted time of TPC-H using ``best-accuracy''. Due to a lack of fine-grained message-size-based modeling, \textbf{Projection I} does not capture the trend of the performance well and is expected to deviate further away as $V$ continues to increase. However, \textbf{Projection I} does reflect how the performance would look if the data communication is less sensitive to message sizes. Another merit is that it also works for the interconnects that we have not studied with microbenchmarks. Although \textbf{Projection II} captures the trend better than the vanilla approach, it consistently underestimates performance due to its ignorance of misalignment. The best approach (\textbf{Projection III}) very accurately models the TPC-H performance. According to it, adding more machines will not improve the performance for $V>6$. 
Figure~\ref{fig:tpch1000-proj-breakdown} shows how the performance breakdown looks when increasing $V$ according to \textbf{Projection III}. As compute time drops with more machines, broadcast time increases due to two factors. First, broadcast becomes less efficient with more machines according to Equation~\ref{eq:broadcast-no-skew}. Second, messages become smaller and cannot efficiently utilize the links. Unlike broadcast, shuffle benefits from more machines (Equation~\ref{eq:shuffle-no-skew}); however, according to our prediction, the shuffle performance will eventually be dragged down by the small message sizes. 

\begin{figure}[t]
     \centering
     \includegraphics[width=0.75\columnwidth]{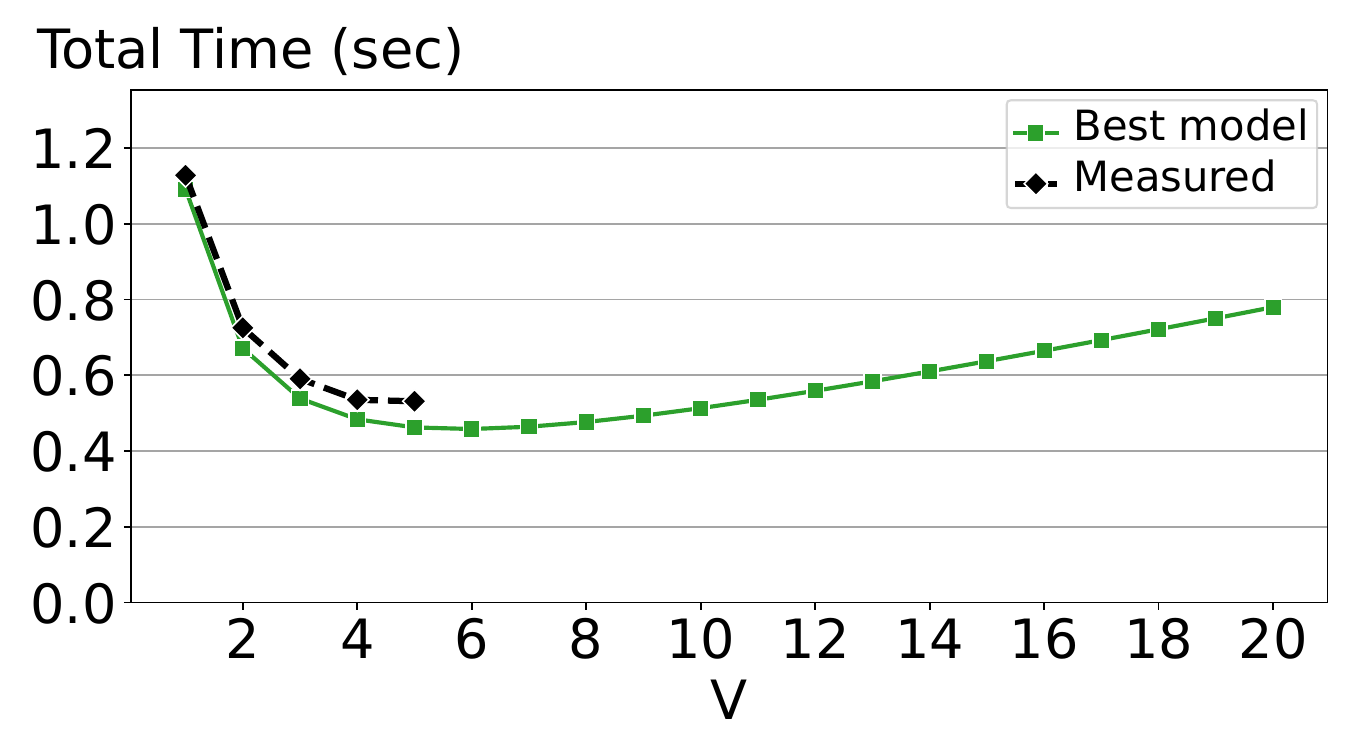}
     \vspace{-2ex}
     \caption{Project TPC-H from $V=1$.\vspace{-5pt}} 
     \label{fig:tpch1000-proj-linear-compute}
\end{figure}

Figure~\ref{fig:tpch1000-proj-linear-compute} shows the projection by the ``best-effort'' (using \textbf{Projection III}). 
It slightly underestimates the performance because the compute cannot scale perfectly linearly, otherwise it provides a very close projection to ``best-accuracy'' (using \textbf{Projection III}). This means even with the performance of $V=1$, we can reliably predict the performance of TPC-H for a large number of machines!


\subsection{Price-Performance}
\label{sec:price-perf}
\begin{figure}[t]
     \centering
     \includegraphics[width=0.75\columnwidth]{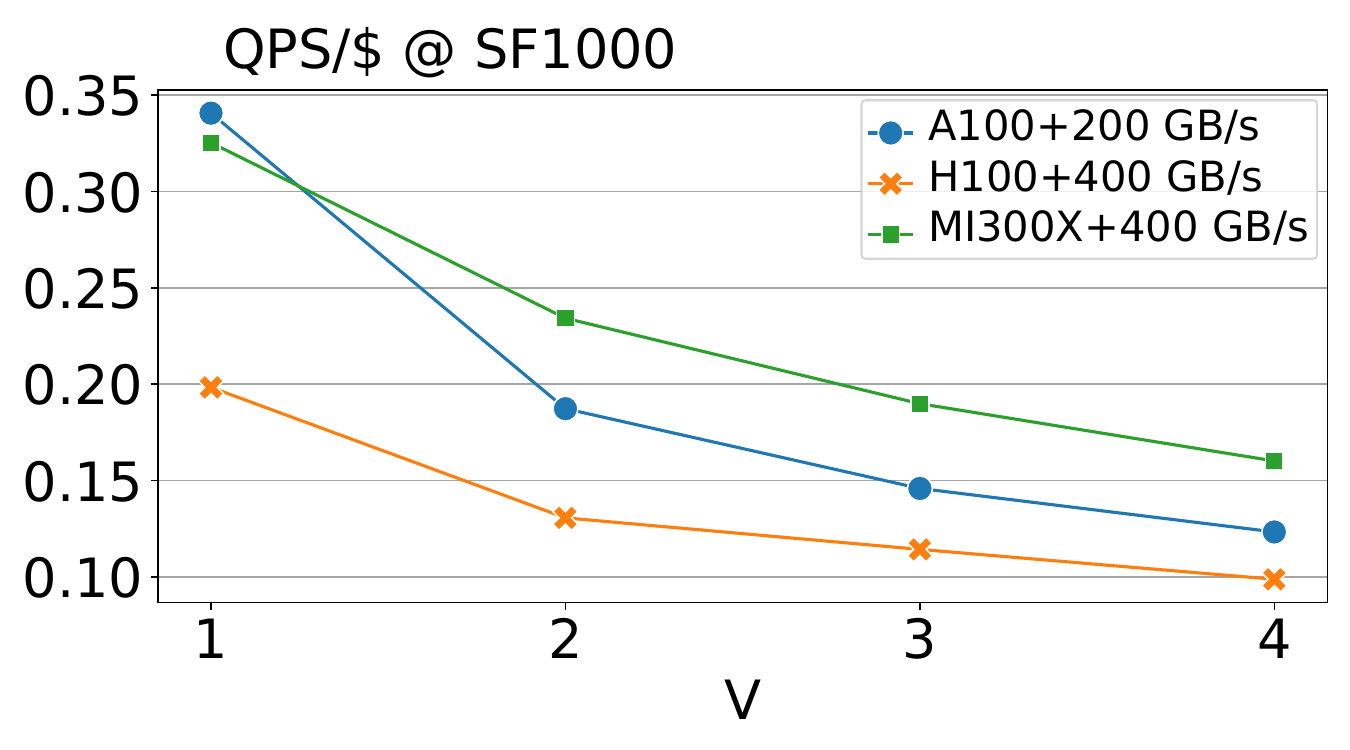}
     \vspace{-2ex}
     \caption{Query-per-second/USD.\vspace{-5pt}}
    \label{fig:price_perf}
\vspace{-2ex}
\end{figure}

Although the workload performance is lower on the A100 VMs compared to that on the H100 and MI300X VMs, they have an advantage from a price-performance perspective. Table~\ref{tab:cluster_config} lists the hourly price of one machine in different clusters. Currently (as of this writing, March 2025), the Azure pay-as-you-go pricing~\cite{azure-vm-pricing} for single eight-GPU VMs is around 3$\times$ and 1.94$\times$ higher for the H100 and MI300X VMs compared to that for the A100 VMs.
Figure~\ref{fig:price_perf} shows the price performance of each cluster when running SF=1000. The price performance is calculated as the ``query-per-second'' (QPS) per dollar. Since our A100 cluster does not have InfiniBand connectivity between VMs, we use the performance projections (vanilla approach) discussed in Section~\ref{sec:perf-projection} to estimate performance for that, assuming 8x200 Gbits/sec InfiniBand NICs per A100 VM.
The result shows that runs for TPC-H SF=1000 on a single eight-GPU A100 VM have 1.72$\times$ and 1.05$\times$ better QPS/\$ respectively compared to the QPS/\$ on the H100 and MI300X VMs. For multi-VM price-performance,  we find that for $V=4$, A100+8x200 Gbits/sec has 1.25$\times$ better and 0.77$\times$ worse QPS/\$ compared to 4 VMs of H100 and MI300X respectively, both of which use 8x400 Gbits/sec InfiniBand NICs per VM.
We can conclude that MI300X VMs are priced competitively, as they have performance similar to that of the H100 VMs, but are currently $\sim$35\% less expensive than them.
\subsection{Message sizes for data exchange}
\label{sec:message-sizes}
\begin{figure}[t]
    \centering
    \begin{subfigure}[b]{0.495\columnwidth}
         \centering
         \includegraphics[width=\columnwidth]{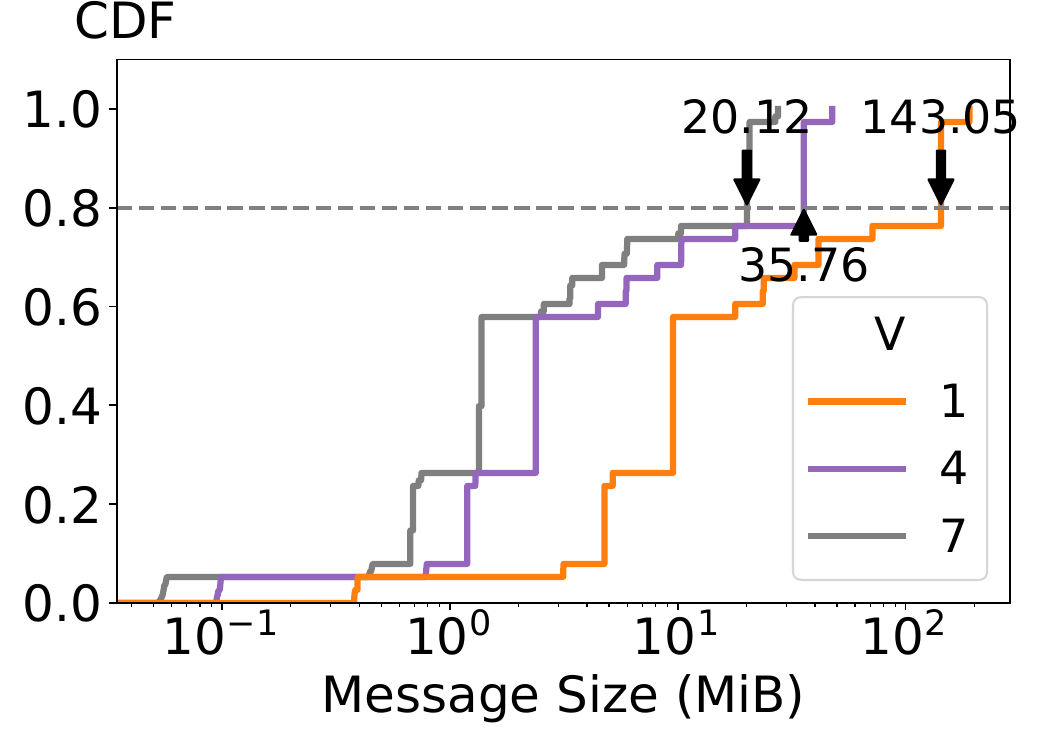}%
         \caption{TPC-H SF=1000. Broadcast}
         \label{fig:ag_msg_size_cdf_tpch1000_a100_eth}
    \end{subfigure}
    \begin{subfigure}[b]{0.495\columnwidth}
         \centering
         \includegraphics[width=\columnwidth]{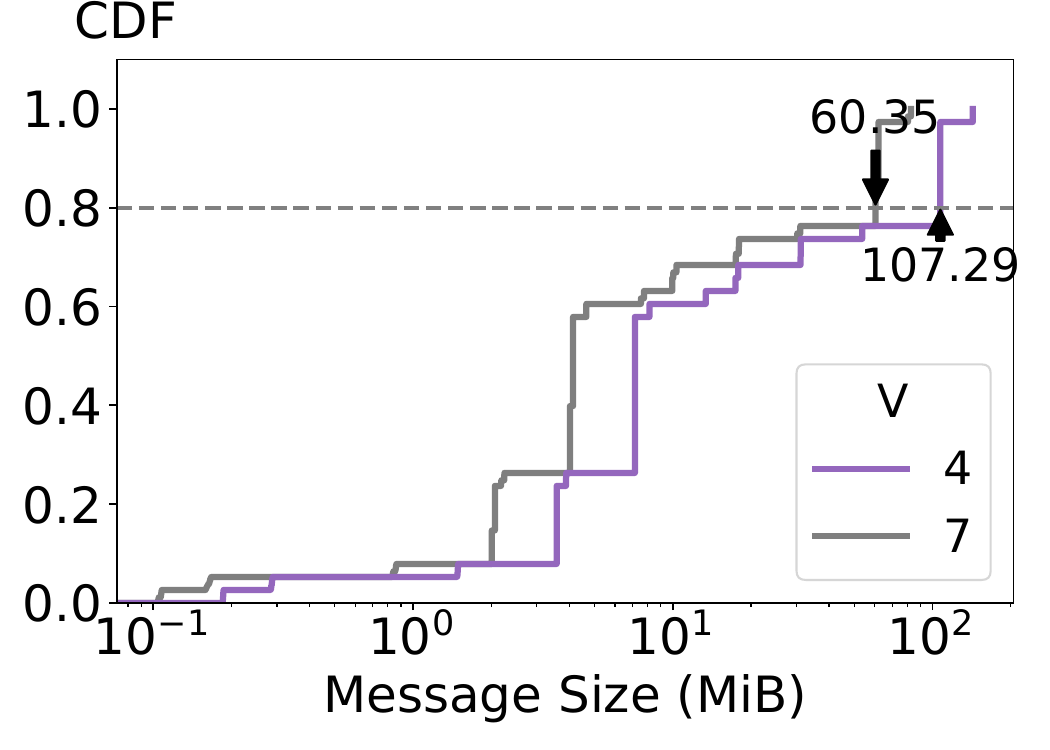}%
         \caption{TPC-H SF=3000. Broadcast}
         \label{fig:ag_msg_size_cdf_tpch3000_a100_eth}
    \end{subfigure}\\
    \begin{subfigure}[b]{0.495\columnwidth}
         \centering
         \includegraphics[width=\columnwidth]{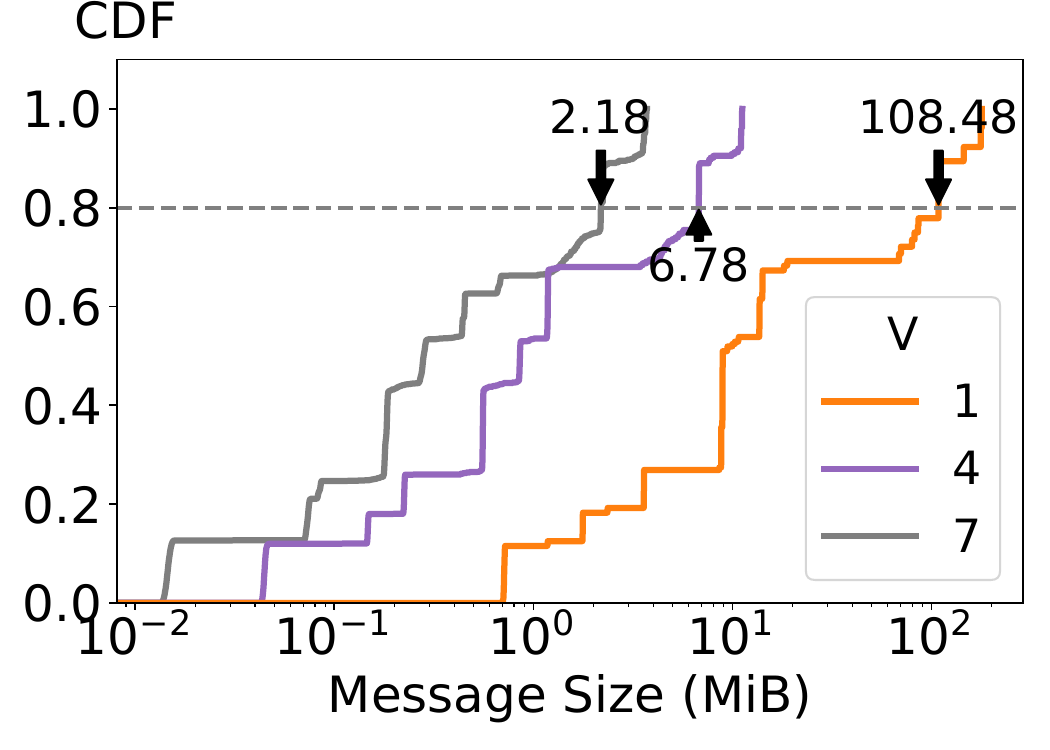}%
         \caption{TPC-H SF=1000. Shuffle.\vspace{-7pt}}
         \label{fig:a2a_msg_size_cdf_tpch1000_a100_eth}
    \end{subfigure}
    \begin{subfigure}[b]{0.495\columnwidth}
         \centering
         \includegraphics[width=\columnwidth]{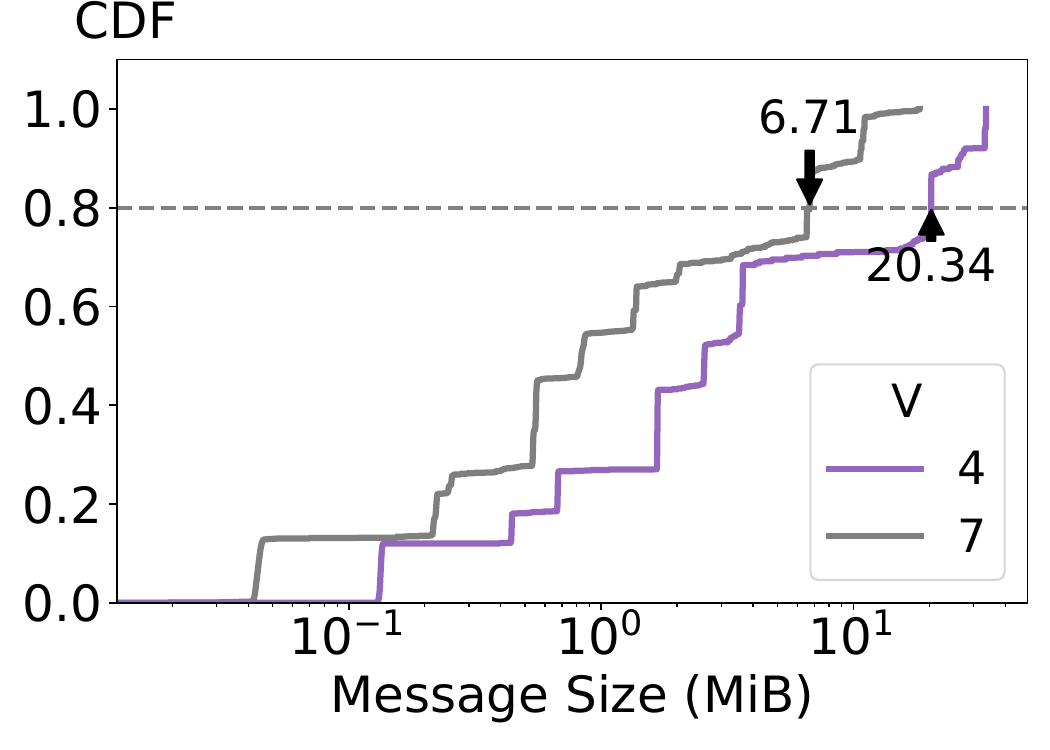}%
         \caption{TPC-H SF=3000. Shuffle.\vspace{-7pt}}
         \label{fig:a2a_msg_size_cdf_tpch3000_a100_eth}
    \end{subfigure}
    \vspace{-3ex}
    \caption{Message size distribution for all running queries. Max. message size decreases as $V$ increases.\vspace{-3pt}}
    \label{fig:msg_size_cdf}
\end{figure}

Figures~\ref{fig:msg_size_cdf} show the distribution of inter-GPU message sizes for broadcasts and shuffles, over all the 22 queries, along with the $80^{\text{th}}$ percentile values. For these workloads, most of the messages are at most a few hundred MiBs and usually much smaller, e.g., for $V=4$, 80\% of shuffles and broadcasts are smaller than 7 MiB and 36 MiB, respectively for SF=1000, and 21 MiB and 108 MiB for SF=3000. The largest message sizes occur for $V=1$. These are 179 MiB and 191 MiB for shuffles and broadcasts at SF=1000. The message sizes increase with dataset size, e.g., up to 537 MiB for shuffles in Q22 at SF=3000 ($V=1$). They also increase with data skew, which we will discuss further in Section~\ref{sec:jcch-analysis}.  

For both message types, increasing $V$ reduces message sizes, but the reduction is larger for shuffles than for broadcasts due to their inverse quadratic scaling with $N=V\times k$ rather than the inverse linear scaling for broadcasts (as we discussed in Section~\ref{sec:models}). Although theoretically, this should result in lower network overheads with increasing $V$, smaller messages utilize the bandwidth less effectively, as we showed in Section~\ref{sec:data-exchange-analysis}, due to protocol and kernel launch overheads, thereby preventing a proportional reduction in per-message network overheads with scale out.

\subsection{Memory Occupancy}
\label{sec:peak-memory}

Compared to main memory in high-end servers, single GPUs have much smaller HBM capacity, thereby limiting the largest dataset size that can be kept resident on each GPU. Using multiple GPUs per machine alleviates, but does not remove, this bottleneck. In our setup, we run queries with all data resident in the GPU HBMs, and do not spill or otherwise transfer data between the GPUs and main/external memory or storage. A query execution will fail if the peak memory consumption exceeds the HBM capacity at runtime. Table~\ref{tab:query-completion-compressed} shows query completion success with different cluster configurations.
While all queries for SF=1000 finish with $V\geq 1$ on any cluster, SF=3000 requires $V\geq 4$ for the A100/H100 clusters, but $V\geq 2$ is sufficient for the MI300X cluster due to the larger HBM capacity. 

\begin{figure}[t]
    \centering
    \begin{subfigure}[b]{0.495\columnwidth}
         \centering
         \includegraphics[width=\columnwidth]{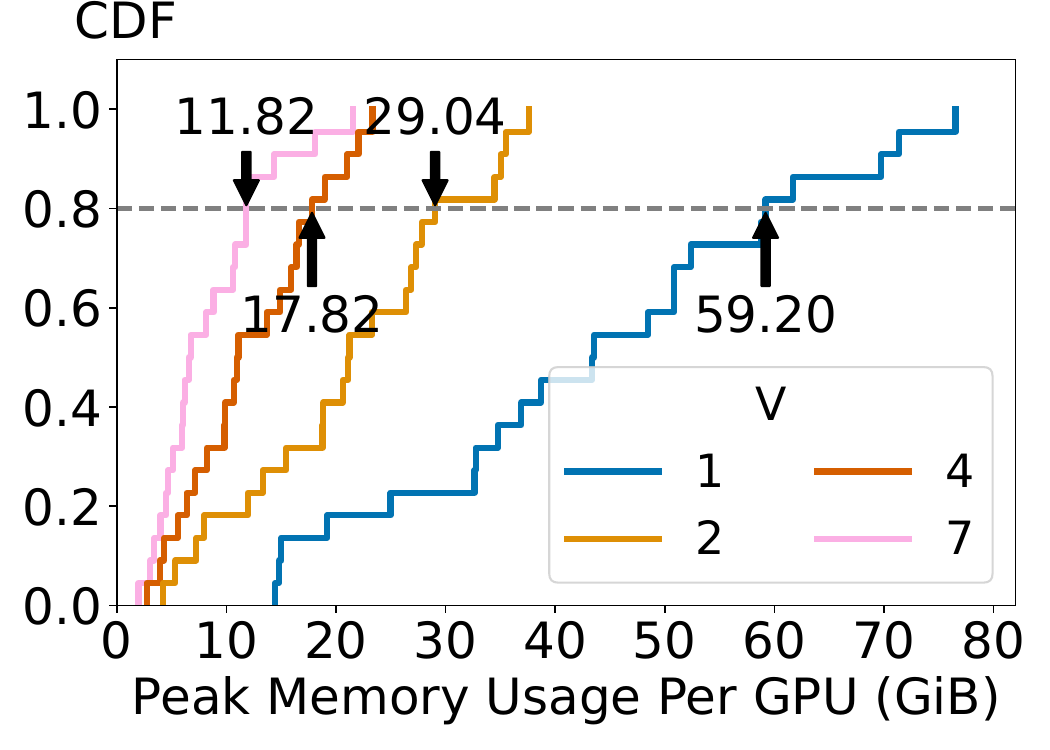}
         \caption{TPC-H SF=1000}
         \label{fig:mem_util_tpch1000_a100_eth}
    \end{subfigure}
    \hfill
    \begin{subfigure}[b]{0.495\columnwidth}
         \centering
         \includegraphics[width=\columnwidth]{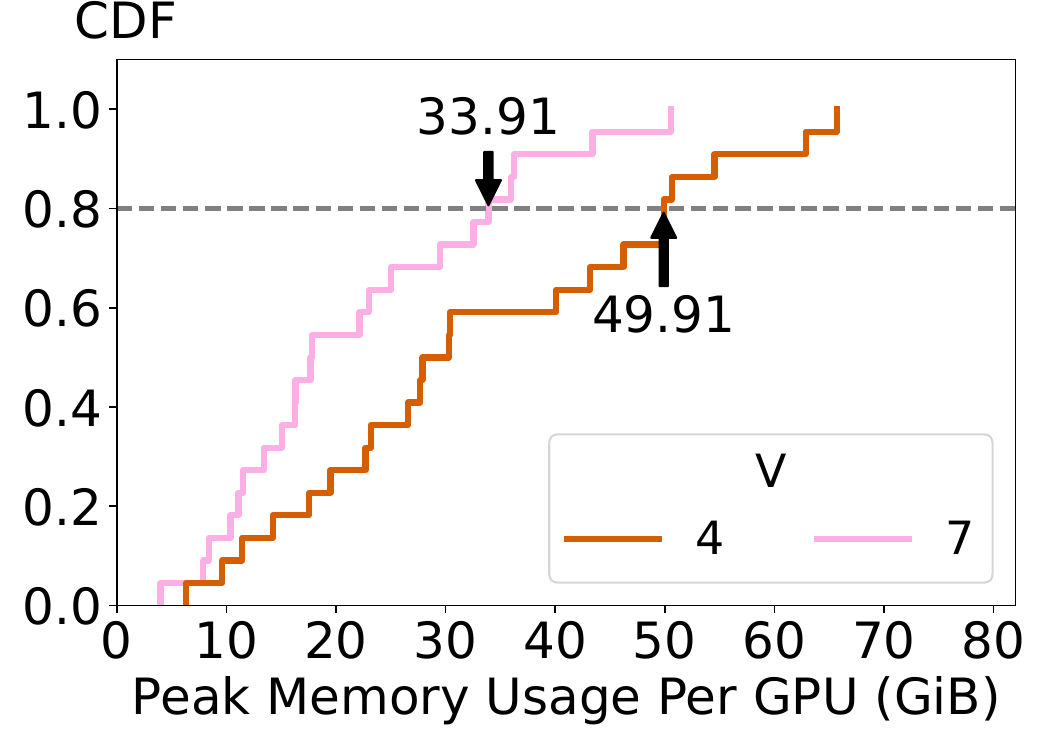}
         \caption{TPC-H SF=3000}
         \label{fig:mem_util_tpch3000_a100_eth}
    \end{subfigure}
    \vspace{-5ex}
    \caption{Distribution of peak GPU memory occupancy of all running queries. Occupancy decreases as $V$ increases.\vspace{-5pt}}
    \label{fig:mem_util_tpch}
\end{figure}
Figure~\ref{fig:mem_util_tpch} shows the distributions, along with the 80$^{\text{th}}$ percentiles, of peak per-GPU HBM occupancy on the A100/H100 GPUs during runs of all the 22 queries. The peak occupancy is affected by the size of the input dataset and intermediate results, and the space needed by GPU kernels and the runtime scheduler. The peak increases with the scale factor, e.g., the 80$^{\text{th}}$ percentile increases from 17.8 GiB for SF=1000 to 49.8 GiB for SF=3000, and decreases with $V$, e.g., the 80$^{\text{th}}$ percentile for SF=1000 decreases from 59.2 GiB for $V=1$ to 11.8 GiB for $V=7$, both due to the associated changes in the per-partition data size for each GPU. 

One difference in memory consumption between single-GPU and multi-GPU query execution is that additional space is needed for the latter to create partitioned inputs for shuffle operations. 
This is not much in our setup since, for this workload, the shuffle message sizes are small (as we discussed in Section~\ref{sec:message-sizes}) and the tables do not have too many columns. Memory needs for partitioning can be reduced by partitioning one column at a time for shuffling and reusing the memory for the next column.

\subsection{Comparison with other Databases}
\label{sec:other_comparison}
In this section, we compare our system with two popular database systems, DuckDB (CPU-based) and HeavyDB (GPU-based). 

We run DuckDB v1.3.2 on a single VM of cluster 2 (Table~\ref{tab:cluster_config}), which has 96 CPU cores and 1900 GiB memory. For each query, we do 10 warmup runs, then measure the median runtime of 10 subsequent runs. For TPC-H SF=1000, DuckDB takes 121 seconds, two orders of magnitude slower than our result for $V=1$ on cluster 2. 
Assuming perfect scaling with $V$, DuckDB would still need at least 24.2 seconds (plus communication overheads) with $V=5$ compared to 0.53 seconds in our case. For TPC-H SF=100, DuckDB spends 10.7 seconds. In contrast, our system spends 0.13 seconds using eight H100 GPUs in a single VM (0.59 seconds using a single H100 GPU). On a 32$\times$-cheaper CPU-only VM with 64 vCPUs (VM type: D64s v5, 32 cores, \$3.072/hr), DuckDB needs 13.1 seconds for SF=100, thus allowing us a performance advantage of over two orders of magnitude, and a QPS/\$ advantage of over 3$\times$ for SF=100 using eight H100 GPUs. 

For HeavyDB\footnote{We were unable to download the enterprise version of HeavyDB, and found that the available version was slower than published numbers.}, we use their published TPC-H performance numbers~\cite{heavydb-perf}. On a GH200 machine, which contains a single H100 GPU, HeavyDB takes 3.9 seconds (without Q21) for SF=100. Assuming perfect scaling, it would need at least 0.49 seconds with eight GPUs, leaving us with a performance advantage of at least 3.75$\times$. 

\section{Performance Sensitivity}
\label{sec:perf-sensitivity}
We now discuss how naive broadcast implementations (Section 7.1), skewed data distributions (Section 7.2), sub-optimally partitioned input data (Section 7.3), and data access from host memory (Section 7.4) can impact the query performance.

\subsection{Collective vs point-to-point broadcasts}
\label{sec:coll-vs-p2p}
\begin{figure}[bt]
    \centering
    \subfloat[H100+Ethernet]{
         \includegraphics[width=0.49\columnwidth]{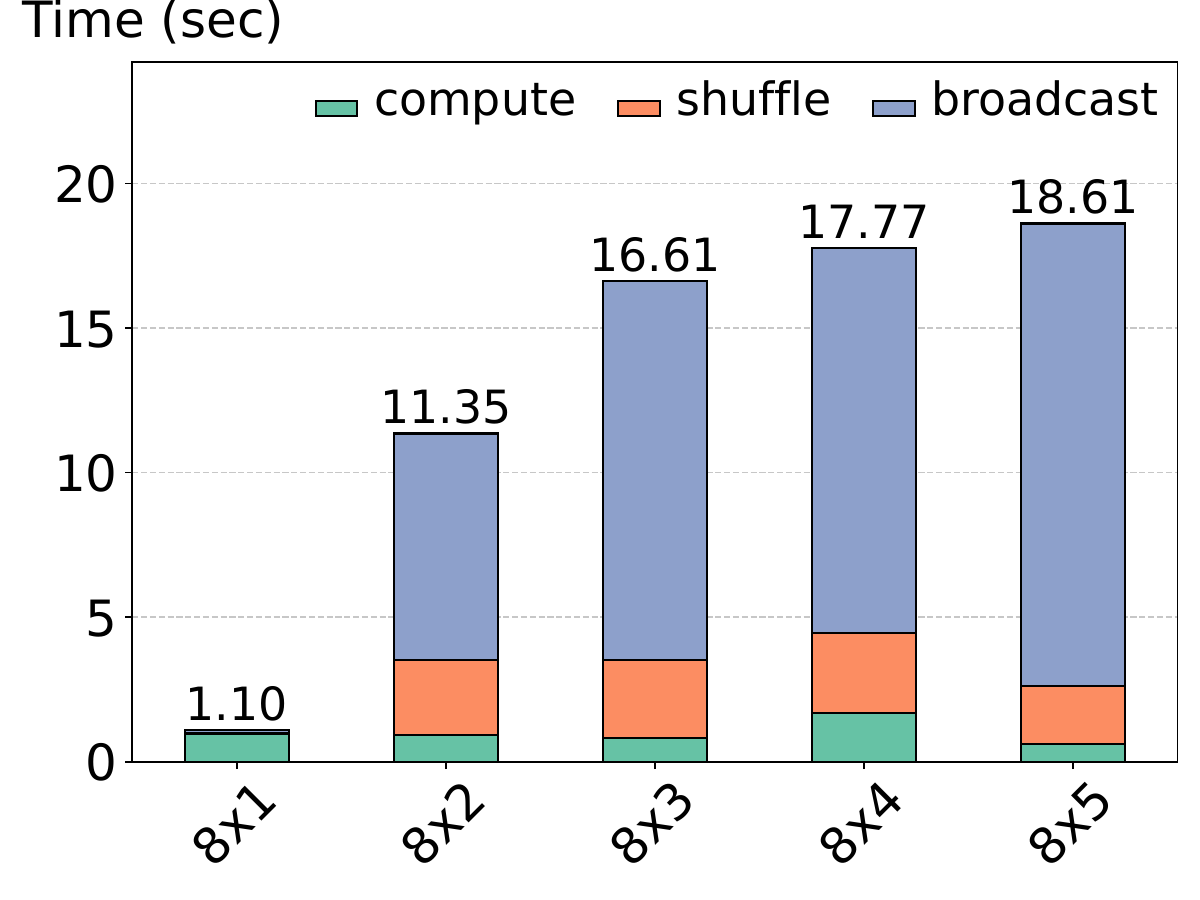}
         \label{fig:total_time_breakdown_tpch1000_h100_eth_p2p}
         \vspace{-5pt}
    }
    \subfloat[H100+InfiniBand]{
         \includegraphics[width=0.49\columnwidth]{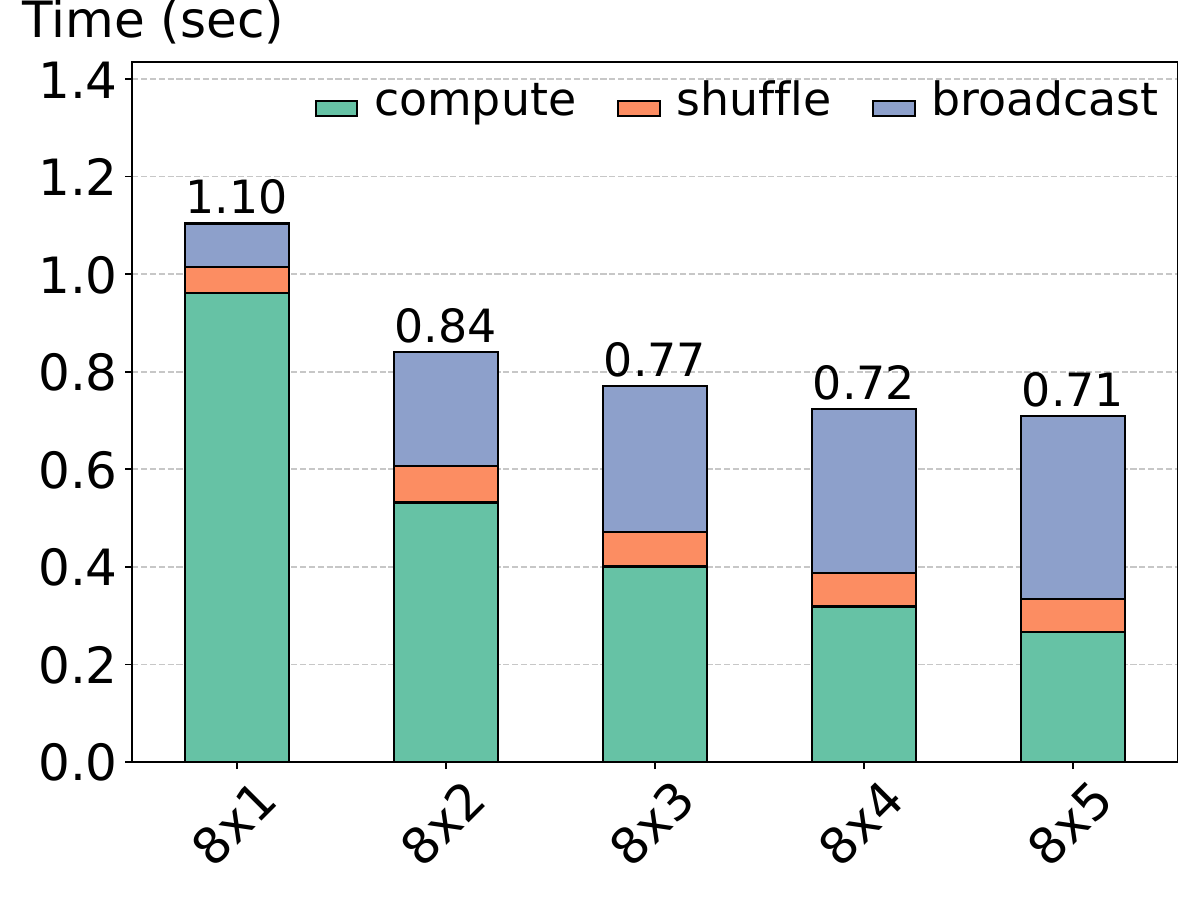}
         \label{fig:total_time_breakdown_tpch1000_h100_ib_p2p}
         \vspace{-5pt}
    }
    \vspace{-2ex}
    \caption{TPC-H SF=1000 using p2p broadcasts.\vspace{-3pt}}
    \label{fig:total_time_breakdown_tpch1000_p2p}
    \vspace{-3ex}
\end{figure}

In Section~\ref{sec:approach}, we emphasized that broadcast needs to be done with a collective operation rather than with a set of point-to-point (p2p) operations for better performance. 
%Here, we detail the impact of p2p-based broadcast on the performance of TPC-H.
Figure~\ref{fig:total_time_breakdown_tpch1000_p2p} shows the impact on broadcast time and workload performance if p2p operations were used. Compared to Figure~\ref{fig:total_time_breakdown_tpch1000} that uses collective operations for broadcasts, the times remain similar for $V=1$. However, the impact increases with $V$, causing increasing slowdowns. For example, for the A100+Eth configuration (not shown in the figure), there is a 1.6$\times$ slowdown for $V=2$, but 2.1$\times$ for $V=7$.The slowdown is 1.6$\times$, 1.2$\times$ at $V=2$, but 2.2$\times$, 1.3$\times$ at $V=5$ for H100+Eth and H100+IB respectively.

The reason for these slowdowns is that with point-to-point operations for broadcast, the same message is unnecessarily transmitted over the network. This is not an issue for intra-VM GPU-GPU high-bandwidth interconnects (e.g., NVLink), but it causes significant slowdowns with lower-bandwidth inter-VM networks. With knowledge of the entire operation, as well as the network topology, collective operations (\texttt{ncclBroadcast}) can reduce or avoid these duplicate transfers and consequent slowdowns.

\subsection{Uniform vs Skewed inputs}
\label{sec:jcch-analysis}

The TPC-H dataset has a largely uniform distribution of data values to table keys~\cite{jcc}. This leads to the partitioned table sizes on the different GPUs being largely similar. To test scenarios involving skewed data, we use the JCC-H dataset with the TPC-H queries. This is feasible since the database schema remains the same, and JCC-H is designed to be a drop-in replacement for TPC-H. 

\begin{figure}[t]
    \centering
    \begin{subfigure}[b]{0.495\columnwidth}
        \centering
        \includegraphics[trim={0ex 3ex 0ex 0ex},clip,width=\columnwidth]{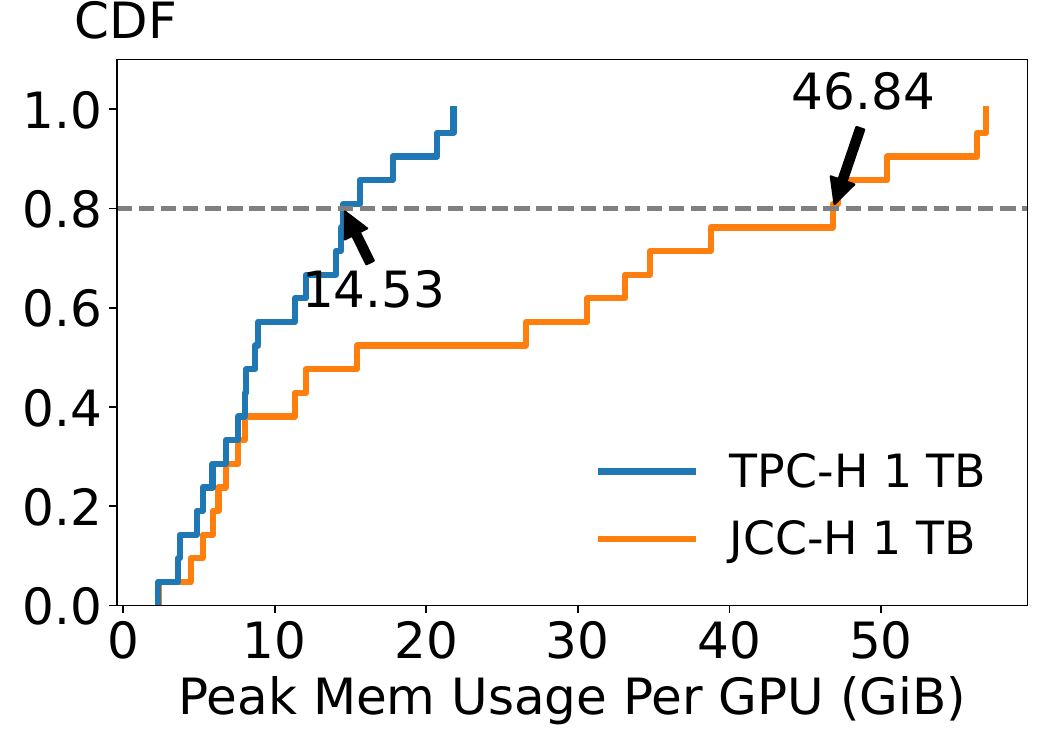}
        \caption{Peak memory utilization CDF.}
        \label{fig:mem_util_tpch_vs_jcch_peak}
    \end{subfigure}%
    \begin{subfigure}[b]{0.495\columnwidth}
        \centering
        \includegraphics[trim={0ex 3ex 0ex 0ex},clip,width=\columnwidth]{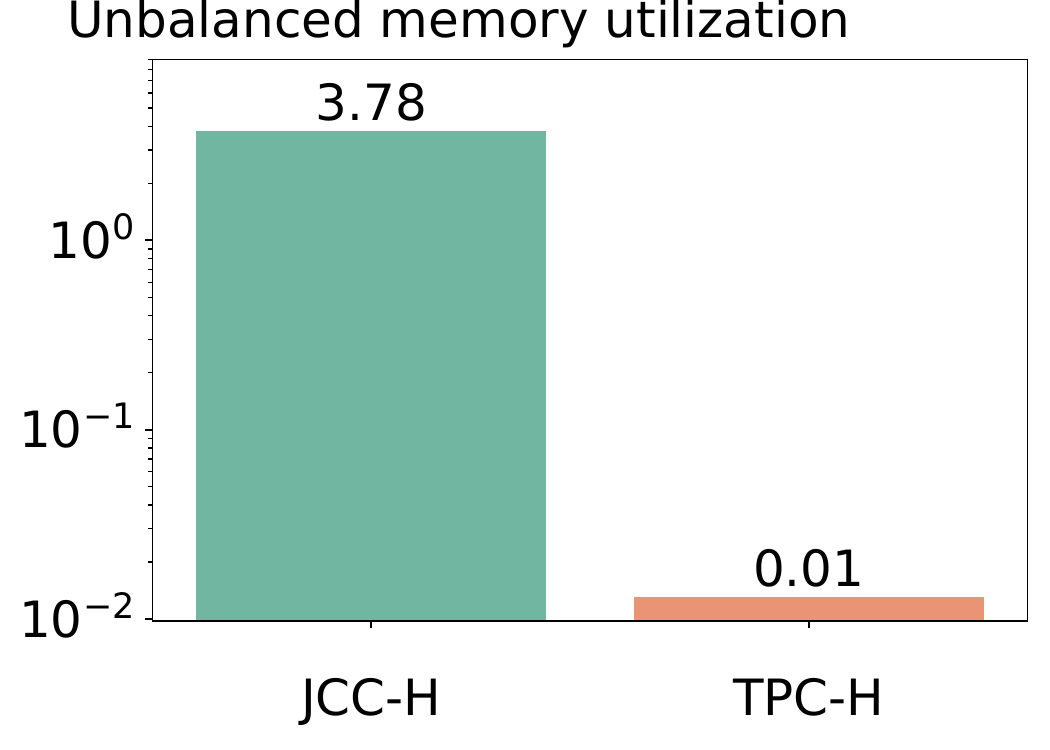}
        \caption{Imbalance across GPUs.}
        \label{fig:mem_util_tpch_vs_jcch_std}
    \end{subfigure}%
    \vspace{-2ex}
    \caption{Memory utilization for TPC-H and JCC-H (1 TB).}
    \label{fig:mem_util_tpch_vs_jcch}
 \end{figure}

\begin{figure}[t]
    \centering
    \includegraphics[width=\columnwidth]{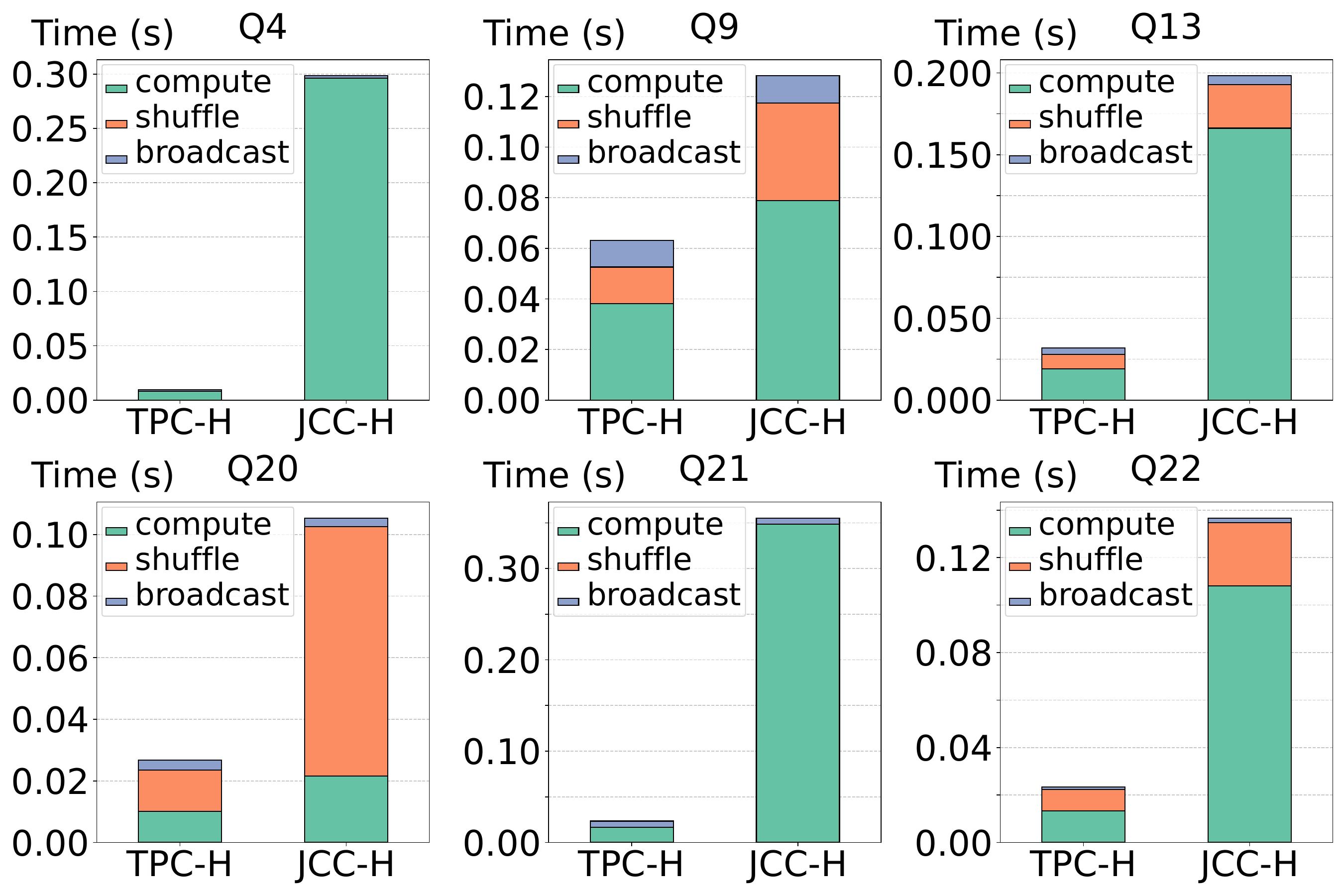}
    \vspace{-5ex}
    \caption{Time breakdown comparison. (V=5)}
    \label{fig:tpch-vs-jcch-breakdown}
    %\vspace{-3ex}
\end{figure}

\noindent\textbf{Memory utilization and message sizes.}
Figure~\ref{fig:mem_util_tpch_vs_jcch_peak} shows the distribution of per-GPU peak memory occupancies of 21 queries\footnote{We exclude Q18 since it currently produces incorrect results for JCC-H.\label{fn:jcc-q18}} for SF=1000 with $V=5$.  
In contrast, we also show the distribution for these 21 queries on TPC-H for $V=5$.
The imbalance caused by the data skew results in more HBM being used on some GPUs, while some other GPUs' HBM remains under-utilized. The 80$^{\text{th}}$ percentile and the max. peak memory used are 46.8 and 56.9 GiB, respectively, for JCC-H, but 14.5 GiB and 21.8 GiB for TPC-H 1TB. In Figure~\ref{fig:mem_util_tpch_vs_jcch_std}, we show the imbalance of memory utilization among GPUs. For each query, we calculate the standard deviation of all GPUs' memory utilization and report the average of all queries. The results indicate that JCC-H causes a severely uneven workload distribution among GPUs. 
A query fails if the memory required exceeds the available HBM capacity on even a single GPU. The A100/H100 clusters need $V\geq 5$ to run these queries, while the MI300X cluster needs $V\geq 3$.
In terms of the message size distribution (not shown), JCC-H 1 TB has a much longer tail than TPC-H 1 TB for the shuffle, whereas the two benchmarks have very similar distributions for the broadcast. The max. message size in broadcast operation for both benchmarks is 38.7 MiB. In contrast, the max. message size in the shuffle operation is 244 MiB and 10 MiB for JCC-H and TPC-H, respectively.

\noindent\textbf{Per-query analysis.}
Several queries severely suffer from the skew introduced in JCC-H, as shown in Figure~\ref{fig:tpch-vs-jcch-breakdown}. The figure details where the time is spent in each benchmark. The two main contributing factors to slower execution are compute and shuffle. 
In Q4, the \texttt{lineitem} table is ill-partitioned based on \texttt{l\_orderkey}, leading to some GPUs processing almost 7$\times$ more data than others. Worse still, due to the imbalance, some GPUs need to build a hash table with around 450M keys, far exceeding the problem size that a hash join can efficiently handle~\cite{wu25-gpu-joins-groupby}. 
In Q20, two GPUs in the same VM need to send around 7$\times$ more data when shuffling \texttt{lineitem} to join with \texttt{partsupp}. Moreover, some GPUs also receive almost 11$\times$ more data than others due to a poor partition function. As discussed in Section~\ref{sec:models} and Section~\ref{sec:data-exchange-analysis}, shuffle can be affected by such a skew, resulting in a sheer increase of shuffle time. Even after shuffling, the \texttt{lineitem} table is still distributed in an unbalanced manner, which causes the subsequent join to be slower. 
Q9 sees a significant increase in both compute and shuffle. The shuffle of \texttt{partsupp} takes longer due to the skew in initial data distribution and the partition function. The compute is slowed down by a series of joins that stress only some GPUs. In contrast, the broadcast is not affected by the skew even though the largest message size in the broadcast is almost 15$\times$ larger than the smallest.

\noindent\textbf{DuckDB comparison.}
We run the same JCC-H benchmark (SF=1000) with DuckDB on a single machine of our cluster 2. Similar to Section~\ref{sec:other_comparison}, we report the median time of warm runs. It takes DuckDB 108.34 seconds to finish 22 queries, 67x slower than our system (1.62 seconds, $V=5$, excluding Q18\footref{fn:jcc-q18}).

\subsection{Partitioned vs non-partitioned inputs}
\label{sec:non-partitioned-impact}

Our performance results so far are for query runs on appropriately partitioned input data that reduces/avoids shuffles on large input tables. Changing the partitioning scheme may require corresponding changes to the query plan for functional correctness and result in a different run time for the query. Here, we investigate its impact using Q12 as an example. This query filters \texttt{lineitem} by applying a predicate and joins with \texttt{orders}. Our default plan does not need any data exchange operations for this query since the input tables (\texttt{lineitem}, \texttt{orders}) are partitioned by their join keys (see Table~\ref{tab:query-stats}). We consider the following alternative query plans when neither table is partitioned on these keys.
\begin{itemize}[leftmargin=*]
    \item  \textbf{Pa}: Shuffle \texttt{lineitem} (filtered) on \texttt{l\_partkey} and \texttt{orders} on \texttt{o\_partkey}, then do the join.
    \item  \textbf{Pb}: Broadcast \texttt{lineitem} (filtered), then join with \texttt{orders}.  
\end{itemize}

\begin{figure}[t]
\centering
\includegraphics[width=0.9\columnwidth]{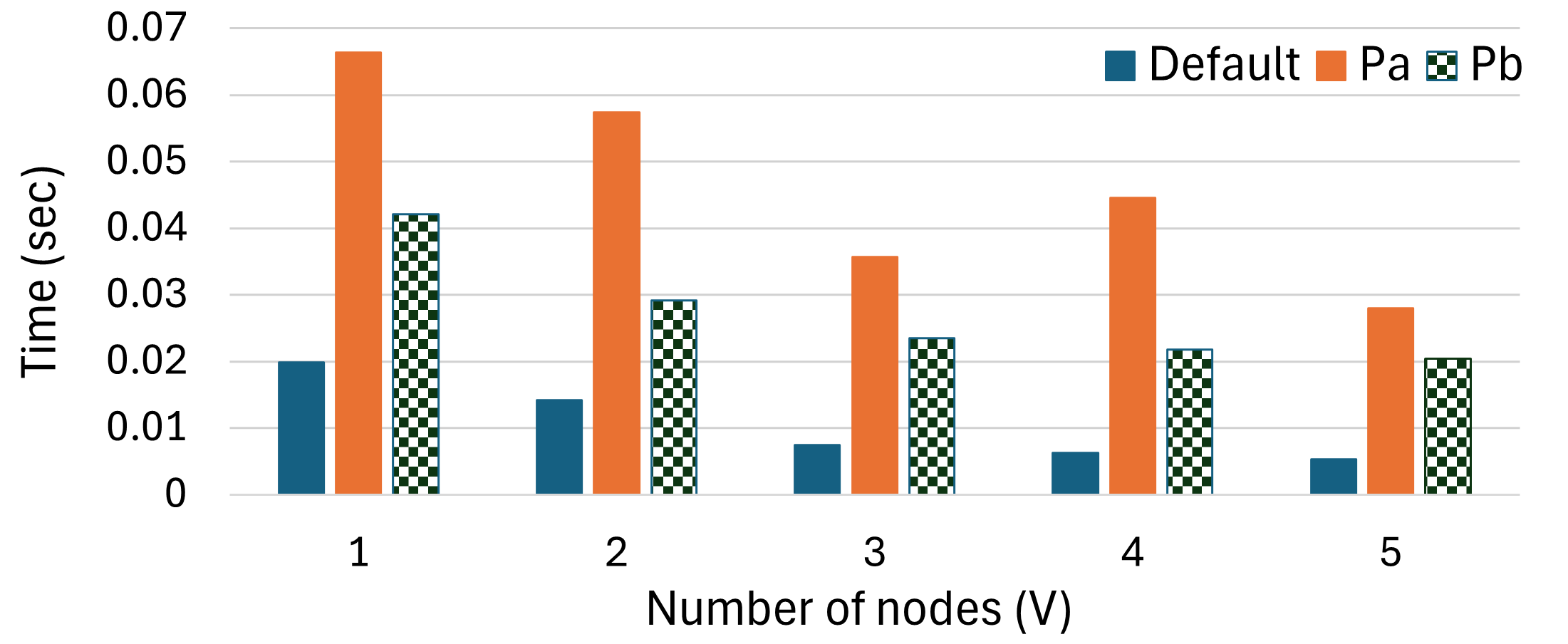}\vspace{-3ex}
\caption{Q12 warm run times for TPC-H 1TB  using different query plans on the H100 cluster with InfiniBand.}
\label{fig:unpartitioned-comparison}
%\vspace{-3ex}
\end{figure}

Figure~\ref{fig:unpartitioned-comparison} shows the run times of Q12 for TPC-H SF=1000 on the H100 cluster with a different number of nodes ($V$). The default plan, which assumes already-partitioned data on join keys and avoids data exchanges, was the best. Plan \textbf{Pa} was the worst, while \textbf{Pb} came second. As already discussed, broadcasts can be more efficient than shuffles, as is the case here. \textbf{Pa} gets more competitive with more nodes due to the faster (quadratic) reduction of message sizes compared to the linear reduction for broadcast (see Section~\ref{sec:models}).  

\subsection{Warm vs cold runs}
\label{sec:cold-runs}

The performance numbers that we presented so far are for warm runs of queries with the input data already loaded in the GPU HBMs. For cold runs, data needs to be moved from the main memory over the CPU-GPU PCIe bus. On the H100 multi-GPU machines, this can be done with an aggregate bandwidth of 440 GB/sec ($8\times55$) over the 8 PCIe gen5 buses (theoretical max. of $8\times63$ GB/sec). The total run time for the 22 queries for TPC-H 1TB using one H100 VM increases from 1.13 secs to 11.4 secs for cold runs, which is still more than 7$\times$ faster than CPU-only warm execution~\cite{HPE_TPC-H_2022}. This is the worst-case scenario, as, in practice, some of the input columns may be cached from prior query executions, thereby not incurring the loading overhead. With multi-machine clusters, the data loading time drops linearly with the number of machines ($V$) assuming that the input data is uniformly partitioned and that the data loads on all machines are initiated in parallel. Other recent architectures~\cite{gh200,gb200} replace the CPU-GPU PCIe with a high-bandwidth interconnect. Thus, we foresee data loading as not being the bottleneck anymore in the future. Additional overheads affecting cold run times are TQP query compilation overheads, which are query dependent ~\cite{surakav_he_2022}, but can be mitigated using a query plan cache. 

\section {Related Work}
\label{sec:related}
\noindent\textbf{Distributed analytical databases.} There is a long history of research and development of distributed databases for data analytics~\cite{polaris,snowflake,bigquery,redshift,databricks,spark,distributed-hyper}. All of these systems only use the CPUs for data processing, and many adopt the Massively Parallel Processing (MPP) paradigm for scalability. Our work adopts MPP to construct a \emph{GPU-based} distributed analytical database. 

\noindent\textbf{Single-GPU query processing.} There are numerous works studying different aspects of single-GPU-based query processing, for example, implementation and optimization~\cite{sioulas19-partitioned-radix-join, wu25-gpu-joins-groupby, Rui17-fastequijoin, kroviakov-crossdevice, deng24-prefetching, Diego18-groupby, Tomas15-groupby, rosenfeld-hash-groupby, Yogatama23-udaf,crystal_shanbhag_2020,hong25-themis,surakav_he_2022,tensor_tea_vldb_2022,gandhi2022tensor,Hu22-tcudb,cudf}, data placement and caching~\cite{Yogatama22-mordred}, fast interconnect technologies~\cite{Lutz20-nonpartitionedjoin, Lutz22-tritonjoin}, optimizing storage I/O~\cite{boeschen24-golap}, data compression~\cite{Shanbhag22-compression}, system integration~\cite{mohr23-boss,Koutsoukos21-modularis,Jungmair22-lingodb}, performance analysis and surveys~\cite{Yuan13-yinyang, Rosenfeld22-revisit, Paul20-revisit-gpujoin, sun23-mmjoin,Cao23-gpudb}, and so on. These techniques are orthogonal to our work but are beneficial in improving the efficiency of individual GPUs in our system.

\noindent\textbf{Multi-GPU query processing.} 
Many works~\cite{Paul21-multigpujoin,maltenberger22-multigpu-sort,thostrup-gpu-rdma-join,guo19-gpudirectrdma-join,gao2021scaling} have studied the implementation of database operators on multiple GPUs. HetExchange~\cite{chrysogelos19-hetexchange} is a database execution model that can exploit parallelism across multi-core CPUs and multi-GPUs. Yuan et al.~\cite{yuan25-vortex} uses multiple GPUs to address the CPU-GPU PCIe bottleneck. Yogatama et al.~\cite{Yogatama25-lancelot} designs a hybrid CPU and multi-GPU query engine. The above work either does not consider a multi-node GPU cluster or does not present full database systems that can complete large-scale TPC-H benchmarks. Our work studies the most general multi-GPU-multi-node case with different GPU models from different vendors and various network technologies.
Some commercial systems~\cite{heavydb,dask-cudf,voltron-data} also support data analytics on multi-GPU systems. Compared to them, we present an in-depth analysis of the TPC-H and JCC-H workloads in addition to end-to-end query runtime. In terms of system design, we advocate the use of ML-style processing with ML-driven high-performance libraries.

\noindent\textbf{Multi-GPU communication.} 
Many previous works~\cite{cowan23-mscclang,kim24-tccl,Cai21-sccl,wang20-blink,Mert24-hiccl,gloo,ucx} provide alternative multi-GPU communication libraries to NCCL and RCCL. 
Weingram et al.~\cite{Weingram23-xccl} compares the state-of-the-art collective communication libraries.
Some works~\cite{Li20-gpu-interconnect,pearson19-cuda-comm,Hidayetoglu24-commbench} evaluate GPU-interconnects and networks. 
Others~\cite{Cho-19-blueconnect,Fei21-sparse-cc,aashaka23-taccl,wang23-topoopt,Lepikhin20-gshard,multigpu-universal-modeling,multigpu-a2a-bruck-algo,flash-a2a} optimize and/or model certain collective communication primitives, such as all-reduce.
Universal Communication X (UCX)~\cite{ucx} and NVSHMEM~\cite{nvshmem} are other interconnect-agnostic frameworks for multi-GPU communications. 
To the best of our knowledge, we are the first to model and optimize the inter-GPU communication for database applications, which are significantly different from ML applications. We are also the first to show how these ML-oriented communication libraries can be used for databases and demonstrate how much time inter-GPU communication takes up in query processing.

\section{Conclusion}
\label{sec:conclude}
We present a distributed implementation of TQP leveraging group communication libraries to process terabyte-scale TPC-H workloads on multi-GPU clusters. 
Our approach allows for seamless portability across different GPU models (A100, H100, and MI300X) and network technologies (Ethernet, Infiniband, NVLink, Infinity Fabric). 
Our experimental evaluation shows that distributed TQP yields very competitive performance results for TPC-H 1 TB and 3 TB scale factors, up to two orders of magnitude faster than public CPU-server results.
To better understand the performance, we develop analytical models for shuffle and broadcast operations, and conduct a detailed analysis of TPC-H workloads. 
We believe that this work unveils the potential of multi-GPU in the SQL analytics domain and provides researchers and practitioners with valuable insights into how to build an efficient multi-GPU database system. 

\clearpage

\bibliographystyle{ACM-Reference-Format}
\balance
%\bibliography{refs}
%%% -*-BibTeX-*-
%%% Do NOT edit. File created by BibTeX with style
%%% ACM-Reference-Format-Journals [18-Jan-2012].

\end{document}